\documentclass[pra,aps,superscriptaddress,twocolumn,floatfix,notitlepage,longbibliography]{revtex4-2}

\usepackage{amsmath}
\usepackage{amsfonts}
\usepackage{amssymb}
\usepackage{mathtools}
\usepackage{graphicx}
\usepackage{color}
\usepackage{xcolor}
\usepackage{xspace}
\usepackage{stix}
\usepackage[sort&compress]{natbib}

\newcommand{\ud}[1]{{#1^{\dagger}}}
\newcommand{\bra}[1]{\left\langle #1\right|}
\newcommand{\ket}[1]{\left| #1\right\rangle}

\newcommand{\Imamoglu}{\u Imamo\=glu\xspace}
\newcommand{\Vuckovic}{Vu\u{c}kovi\'c\xspace}

 \newcommand{\mean}[1]{\langle#1\rangle}

\usepackage[colorlinks, linkcolor=magenta, citecolor=magenta,
urlcolor=blue]{hyperref}
\begin{document} 
\title{Loss of Antibunching}

\author{Juan Camilo {L\'opez Carre{\~n}o}}
\affiliation{Faculty of Science and Engineering, University of
  Wolverhampton, Wulfruna St, Wolverhampton WV1 1LY, UK}
\author{Eduardo {Zubizarreta Casalengua}}
\affiliation{Faculty of Science and Engineering, University of Wolverhampton, Wulfruna St, Wolverhampton WV1 1LY, UK}
\author{Blanca Silva}
\affiliation{Departamento de F\'isica Te\'orica de la Materia Condensada,
  Universidad Aut\'onoma de Madrid, 28049 Madrid, Spain}
\author{Elena del Valle}
\affiliation{Faculty of Science and Engineering, University of Wolverhampton, Wulfruna St, Wolverhampton WV1 1LY, UK}
\affiliation{Departamento de F\'isica Te\'orica de la Materia Condensada,
  Universidad Aut\'onoma de Madrid, 28049 Madrid, Spain}
\author{Fabrice P. Laussy} 
\affiliation{Faculty of Science and Engineering, University of Wolverhampton, Wulfruna St, Wolverhampton WV1 1LY, UK}
\affiliation{Russian Quantum Center, Novaya 100, 143025 Skolkovo, Moscow Region, Russia}

\date{\today}

\begin{abstract}
  We describe some of the main external mechanisms that lead to a loss
  of antibunching, i.e., that spoil the character of a given quantum
  light to deliver its photons separated the ones from the
  others. Namely, we consider contamination by noise, a time jitter in
  the photon detection and the effect of frequency filtering (or
  detection with finite bandwidth). The emission from a two-level
  system under both incoherent and coherent driving is taken as a
  particular case of special interest. The coherent case is further
  separated into its vanishing (Heitler) and high (Mollow) driving
  regimes. We provide analytical solutions which, in the case of
  filtering, unveil an unsuspected structure in the transitions from
  perfect antibunching to thermal (incoherent case) or uncorrelated
  (coherent case) emission. The experimental observations of these
  basic and fundamental transitions would provide another compelling
  evidence of the correctness and importance of the theory of
  frequency-resolved photon correlations.
\end{abstract}

\maketitle

\section{Introduction}

Antibunching~\cite{paul82a} describes one of the most popular type of
quantum light, the one for which photons get separated from each other
and avoid the opposite bunching tendency of bosons to appear clumped
together~\cite{hanburybrown56a}. Through its observation by Kimble in
1977~\cite{kimble77a}, it provided the first direct evidence of
quantization of the light field, that is to say, the first
observation, albeit indirect, of photons. 
Antibunched light is also of considerable importance for quantum
applications, for instance to feed quantum gates or for the already
commercialized quantum cryptography, in which case one seeks the
ultimate antibunching where transform-limited
photons~\cite{kuhlmann15a} are never detected more than one at a
time. This would provide the ultimate antibunching. This is an
asymptotic race, however, as perfect antibunching has still not been
achieved and some residual multiple-photon emission has always
accompanied the most crafted setups. In principle, since we are
dealing with a quantized property, with a gap separating its value
(one photon) from its neighbours (vacuum and two or more photons),
there is no a priori reason why one could not observe perfect
antibunching, just as one does observe perfect conductivity from a
superconductor or perfect flow from a superfluid. In all these cases,
there are experimental limitations, inaccuracy of measurements, finite
times and energy involved, and yet, one can show that the measurement
of resistance is compatible with a mathematical zero in a
superconductor, where experiments with superconducting coils have
demonstrated current flow persisting for years without degradation,
points to a current lifetime of at least 100,000 years and with
theoretical estimates for the lifetime of a persistent current to
exceed the lifetime of the universe~\cite{gallop_book90a}. This is in
this sense that one can speak of the resistance becoming ``truly
zero'' in a superconductor. Instead, when it comes to the more basic
problem of detecting a single photon, one finds instead deviation from
uncorrelated light of at best (to the best of our
knowledge)~$7.5\times10^{-5}$~\cite{schweickert18a}
and~$9.5\times10^{-5}$~\cite{hanschke18a} which are, furthermore,
sensibly better than most values reported in an ample literature
(which cannot be browsed completely even if we narrow it down to
recent reports below
$10^{-2}$~\cite{muller14a,wei14a,sapienza15a,miyazawa16a,ding16a,somaschi16a,unsleber16a,wang16a,higginbottom16a,huber17a,schweickert18a,senellart17a,hanschke18a,liu18b,musial20a},
see Ref.~\cite{arakawa20a} for a recent review.)  The record-value
antibunching~\cite{schweickert18a, hanschke18a} were significantly
improved by counteracting re-excitations of the two-level system by
implementing improved two-photon excitation schemes (In
Ref.~\cite{lopezcarreno16b}, we have also discussed how exciting a 2LS
with quantum light indeed improves its single-photon
characteristics). But even with this newly added trick to suppress
multiphoton emission, the perfect antibunching of an exact zero---or
no-coincidence at all however long the experiment is run---is still
out of reach.  In this text, we discuss mechanisms that leads to a
loss of antibunching, regardless of the source of light itself which
can, indeed, be perfectly antibunched. We cover both technical (noise,
time-jitter) as well as fundamental reasons (linked to
photon-detection).

\section{Definition of antibunching}

The definition of antibunching requires some discussion, as it varies
throughout times and authors~\cite{teich88a} and is commonly confused
for another one (subpoissonian statistics).  At the heart of every
definition one finds Glauber's theory of optical
coherence~\cite{glauber63c}, that introduces correlation
functions~$g^{(n)}$ for the $n^\mathrm{th}$-order coherence as
\begin{multline}
  \label{eq:Tue6Jun133237BST2017}
  g^{(n)}(t_1,\cdots,t_n)\equiv\\ \frac{\mean{\ud{a}(t_1) \ud{a}(t_2)
      \cdots \ud{a}(t_n) a(t_n) \cdots a(t_2) a(t_1)}}{
    \mean{\ud{a}(t_1)a(t_1)} \mean{\ud{a}(t_2)a(t_2)}\cdots
\mean{\ud{a}(t_n)a(t_n)} }\,,
\end{multline}
where we have assumed that the times~$t_k$ are in increasing order,
i.e.,~$t_1 < t_2 < \cdots < t_n$, and we have assumed a single bosonic
mode~$a$, which is the best way to root our discussion at its most
fundamental level, since involving a continuum from the start should
eventually bring us to the same results. At the two photons level,
that of more common occurence, Eq.~(\ref{eq:Tue6Jun133237BST2017})
reads
\begin{equation}
  \label{eq:Sat31Jul141755CEST2021}
  g^{(2)}(t_1,t_2)=
  {\langle\ud{a}(t_1)\ud{a}(t_2)a(t_2)a(t_1)\rangle\over\langle(\ud{a}a)(t_1)\rangle\langle(\ud{a}a)(t_1)\rangle}
\end{equation}
and if dealing with a steady state, so that only the time
difference~$\tau\equiv t_2-t_1$ matters:
\begin{equation}
  \label{eq:Tue26May162732CEST2020}
  g^{(2)}(\tau)=
  {\langle\ud{a}\ud{a}(\tau)a(\tau)a\rangle\over\langle\ud{a}a\rangle^2}\,.
\end{equation}
At zero time delay~$\tau=0$, Eq.~(\ref{eq:Tue26May162732CEST2020})
further simplifies to
\begin{equation}
  \label{eq:Tue26May162850CEST2020}
  g^{(2)}(0)={\langle{a^{\dagger 2}}a^2\rangle\over\langle\ud{a}a\rangle^2}\,,
\end{equation}
and the application of Eq.~(\ref{eq:Tue26May162850CEST2020}) on a
density matrix~$\rho$ yields the two-photon coincidences $g^{(2)}(0)$
in terms of the probabilities~$p(n)\equiv\bra{n}\rho\ket{n}$ as:
\begin{equation}
  \label{eq:Tue26May162937CEST2020}
  g^{(2)}(0)={\sum_{n=0}^\infty n(n-1)p(n)\over(\sum_{n=0}^\infty np(n))^2}\,.
\end{equation}

Antibunching is nowadays commonly defined by the
condition~\cite{mandel_book95a, scully_book02a}:
\begin{equation}
  \label{eq:Tue6Jun133555BST2017}
  g^{(2)}(0)<g^{(2)}(\tau) \hbox{ for all $0\le\tau<\tau_\mathrm{max}$}\,,
\end{equation}
where~$\tau_\mathrm{max}$ can be infinite. Since for time delays long
enough, photons are uncorrelated, which means, by definition, that the
numerator in Eq.~(\ref{eq:Tue26May162732CEST2020}) factorizes in the
form of the denominator, i.e.,
$\lim_{\tau\rightarrow\infty}g^{(2)}(\tau)=1$, then a popular
understanding of antibunching reads
\begin{equation}
  \label{eq:Tue26May164854CEST2020}
  g^{(2)}(0)<1\,.
\end{equation}
This is inaccurate at best since Eqs.~(\ref{eq:Tue6Jun133555BST2017})
and~(\ref{eq:Tue26May164854CEST2020}) are logically independent, i.e.,
none implies the other, although they are strongly related to each
others~\cite{teich83a}. This point has been made by Zou and
Mandel~\cite{zou90a}. A proper name for
Eq.~(\ref{eq:Tue26May164854CEST2020}) is sub-Poisson
light~\cite{teich88a} (other denominations can be found such as
``photon-number-squeezed light''~\cite{teich88a}). Such discussions
truely become important for particular, and often, odd cases such as
antibunching of super-Poissonian light. In this text, we will focus on
the simplest case that is also that of greatest (today's) interest of
sub-Poissonian antibunched light, so such precautions in the
terminology will not be entirely necessary. By antibunching, we shall
thus understand the tendency of emitting single photons, as is often
the case in the literature anyway. Note that our formalisms and
results could nevertheless be applied to all types of photon
correlations (superbunching, etc.) but we will focus presently on
antibunching.

\section{Measurement of antibunching}

The typical setup for measuring antibunching experimentally is that
designed by Hanbury Brown~\cite{hanburybrown52a} to implement an
intensity interferometer following his naked-eye observation of radar
correlations in the early days of its elaboration. While initially
designed for interferometry in radio-astronomy~\cite{hanburybrown54a},
its application to visible light was quickly understood as involving
photon correlations at the single-particle level, which initially
caused much controversy but was quickly confirmed
experimentally~\cite{hanburybrown56b} (the denomination of
``coherent'' for the beams of light in Ref.~\cite{hanburybrown56b}
predates Glauber's theory of optical coherence and refers to
monochromatic thermal light). The theory of the effect by Twiss gives
to the setup its famed name of Hanbury Brown--Twiss (HBT)
interferometer. While designed for bunching---the natural tendency of
bosons which symmetric wavefunctions tend to clutter together---the
same setup is apt to measure all types of photon correlations,
including antibunching, as had been readily
predicted~\cite{purcell56a,stoler74a}.  The HBT setup consists in a
beam-splitter followed by two detectors in each branch which are
temporally correlated. In practice, the first detector that records a
click starts a time-counter while the other detector stops it and a
normalized histogram of the time differences~$\tau$ between the
successive photons thus reconstructs the second-order coherence
function~$g^{(2)}(\tau)$.  It is also known as a photon-coincidence
measurement. The critical elements that affect the quantum
correlations in this setup are the detectors, which are typically APDs
(Avalanche photodiodes)~\cite{silberhorn07a}. The most frequently used
detectors for detection of low-intensity light are photomultiplier
tubes (PMTs), but their quantum efficiency is small (smaller than
50\%). For this reason APDs are used, which have an additional gain
mechanism, the `avalanch effect'. With the APDs, a stable gain on the
order of $10^2$ to $10^3$ can be achieved, which is still too low to
detect single photons. For this purpose, the APDs must be used in the
`Geiger mode'~\cite{kardynal08a}. These single-photon avalanche
photodiodes (SPADs) have a high detection efficiency and low dark
count rates, but they are slow and with a big timming jitter
(typically $300-400$~ps, with low values of
$35$~ps~\cite{hadfield09a}). To multiply the signal, they use
semiconductor materials. Depending on these materials, the APDs
operate in different frequency windows between $550$ and
$1550$~nm. Another source of noise characteristic of APDs is the
afterpulsing, which can limit the count rate~\cite{yuan07a}.

New methods have emerged to measure photon correlations, in particular
one that relies on a direct observation of the photon streams as
measured by a streak camera~\cite{wiersig09a}. The detected photons
are first transformed into photoelectrons by means of a
photocathode. These new photoelectrons are deflected vertically to
different pixels on the detector as a function of time. Thanks to this
shift, the vertical position on the detector defines the time of
arrival of the photon.  Streak cameras have low detection efficiency
but allow for a resolution of the order of the ps. They operate in
frequency windows of $300-1700$~nm, depending on the material used for
the photocathode. One advantage of a streak camera setup in a cw
regime is that it provides the raw result with no need for
post-processing or normalization. Namely, the
condition~$g^{(2)}(\tau\to\infty)=1$ which is used to normalize the
signal in the case of an HBT measurement, should be automatically
verified with the streak setup. Failure to be the case should point at
some problem in the detection, e.g., non-stationarity of the
signal~\cite{blancathesis}. This also allows to compute higher-order
photon correlations, which can also be achieved with other emerging
techniques such as transition edge sensors set up to direclty resolve
the number of detected photons~\cite{klaas18a}.

With this brief overview of some of the main methods to measure
antibunching, one gets a feeling of the mechanisms that lead to its
loss and that we will model theoretically in the following. These
include, basically, external noise and time uncertainty in the
detection. The latter can be due to a jitter, meaning that a
fluctuation or scrambling of the arrival time due to the detector, or
at a more fundamental level, be linked to the time-energy uncertainty
which is inherent even to ideal detectors. We will cover both
mechanisms. Interestingly, photon-losses, which constitute an
important limitation of all schemes of photon detections, are not
detrimental for the measurement of antibunching. This merely dims the
signal, but preserves its statistics. Although only a coherent signal
can pass linear optical element without being distorted, and that
subpoissonian or superpoissonian signals get closer to Poissonian
distributions, e.g., by passing through
beam-splitters~\cite{bachor_book04a}, this, however, refers to the
noise of the signal rather than to its statistics. A well ordered
stream of single photons would appear less ordered in the presence of
losses, but this would in no way lead to spurious coincidences.  Such
a closeness to a Poissonian distribution is typically measured by the
Fano factor, which relates the width of the input distribution to the
expected one for a Poissonian distribution with the same average
number as
$F = (\langle n^2\rangle - \langle n\rangle ^2)/\langle n \rangle
$. Therefore, as the Fano factor can grow linearly from 0,
corresponding to a Number-state distribution, to 1, corresponding to a
Poisson distribution, as the probability to loose any one photon of
the input beam is increased, the statistics as measured by
$g^{(2)}(0)$ remains constant, equal to that of the ideal signal. This
is clear on physical grounds since removing photons to an antibunched
signal cannot create bunching. What is spoiled is the signal, which
can however be compensated by longer integration times. The loss is
therefore in quantity, not in quality.  This features makes the lossy
setups, including the HBT one, able to measure antibunching as it does
not matter that strictly successive photons be recorded,
$g^{(2)}(\tau)$ being a density probability for any two photons to be
separated by the interval~$\tau$, regardless of whether other photons
are present in between or not. In fact, a histogram of exactly
successive photons would fail to produce the uncorrelated plateau at
long~$\tau$. We can therefore already strike out one of the main
difficulties encountered in the experiment and focus on the other
above-cited mechanisms. Before turning to them in detail, we first
review the antibunching from the source we shall use to illustrate the
general theory, which is of great interest regardless, being the most
fundamental and widespread type of single-photon source.

\section{Examples of antibunching}

The two-level system (2LS) is the paradigmatic source of single
photons. When the emission occurs with the system relaxing from its
excited to its ground state, and since it takes a finite amount of
time for the 2LS to be re-excited, together with the impossibility to
host more than one excitation at a time, two photons can never be
emitted simultaneously.  This is at least the basic picture which one
can form and that applies in the simplest cases of incoherent
excitation as well as strong coherent excitation. Under weak coherent
excitation, on the other hand, subtle interferences at the
multi-photon level also produce antibunching but with a distinct
physical origin~\cite{hanschke20a}. In the remaining of the text we
will work with these three cases: i) incoherent and ii) coherent
excitation, the latter being further separated into its weak (Heitler)
and high (Mollow) regimes.

\subsection{Incoherent excitation}

The Hamiltonian of an incoherently driven 2LS is simply its free
energy, namely
\begin{equation}
  \label{eq:Fri8May2020195906BST}
  H_\sigma = \omega_\sigma \ud{\sigma}\sigma\,,
\end{equation}
where the 2LS is described through the annihilation operator~$\sigma$,
which satisfies the algebra of pseudo-spins, and~$\omega_\sigma$ is
the natural frequency of the 2LS. Both the excitation and decay of the
2LS is taken into account by turning to a master equation (we use~$\hbar=1$ along the paper)
\begin{equation}
  \label{eq:Mon11May2020131426BST}
  \partial_t \rho = i[\rho,H] + \sum_k \mathcal{L}_{c_k}\rho\,,
\end{equation}
with the Lindblad
terms~$\mathcal{L}_\ud{\sigma} \rho = (P_\sigma/2) (2\ud{\sigma}\rho
\sigma - \sigma\ud{\sigma} \rho - \rho \sigma \ud{\sigma})$,
where~$P_\sigma$ is the rate of excitation and
$\mathcal{L}{\sigma} \rho = (\gamma_\sigma/2) (2{\sigma}\rho
\ud{\sigma} - \ud{\sigma}{\sigma} \rho - \rho \ud{\sigma}{\sigma})$,
where~$\gamma_\sigma$ is the decay rate. It is then a simple algebraic
procedure to obtain the second-order correlations of a 2LS under
incoherent driving as
\begin{equation}
  \label{eq:Thu23Apr2020225637BST}
  g^{(2)}_{\sigma,P_\sigma} (\tau) = 1-e^{-\Gamma_\sigma \tau}\,,
\end{equation}
in terms of the effective decay
rate~$\Gamma_\sigma\equiv\gamma_\sigma + P_\sigma$ that causes the
power-broadening of the emission spectrum.

\subsection{Coherent excitation}

The counterpart of the previous section but for coherent excitation is
described by supplementing the master
equation~(\ref{eq:Mon11May2020131426BST}) with the Hamiltonian
\begin{equation}
  \label{eq:Mon11May2020131538BST}
  H_\sigma = \omega_\sigma \ud{\sigma}\sigma + \Omega_\sigma (\sigma
  e^{i\omega_\mathrm{L} t} + \ud{\sigma} e^{-i \omega_\mathrm{L}t})\,,
\end{equation}
where~$\Omega_\sigma$ is the intensity at which the 2LS is driven by a
laser, which emits photons at a frequency~$\omega_\mathrm{L}$. The
temporal dependence of the Hamiltonian in
Eq.~(\ref{eq:Mon11May2020131538BST}) is removed by making a Dirac
transformation, leaving the contribution from the free energy of the
2LS and of the detector proportional to the \emph{detuning} between
the laser and their natural frequencies, namely,~$\Delta_c \equiv
(\omega_c -\omega_\mathrm{L})$ for~$c=\sigma,a$.

The same techniques applied to this variation of the problem provide
the correlations for the 2LS under coherent excitation as
\begin{equation}
  \label{eq:Thu23Apr2020230012BST}
  g^{(2)}_{\sigma,\Omega}(\tau) = 1 -e^{-3\gamma_\sigma\tau/4}
  \left[\cos\left(\frac{\gamma_\mathrm{M}\tau}{4} \right) +
    \frac{3\gamma_\sigma}{\gamma_\mathrm{M}} \sin
    \left(\frac{\gamma_\mathrm{M}\tau}{4}\right)  \right]\,,
\end{equation}
where
$\gamma_\mathrm{M}\equiv\sqrt{\gamma_\sigma^2-(8\Omega_\sigma)^2}$
(`M' is for ``Mollow'') and~$\Omega_\sigma$ is the intensity of the
driving laser.  This more involved expression follows from the two
regimes of low and high driving.  In the Heitler
regime~\cite{heitler_book44a}, where the rate of excitation is much
weaker than the decay rate of the 2LS, the correlations simplify to
\begin{equation}
  \label{eq:Thu23Apr2020230640BST}
  g_{\sigma,\Omega\rightarrow0}^{(2)}(\tau) =
  \left(1-e^{-\gamma_\sigma \tau/2} \right)^2\,,
\end{equation}
and in the limit of large driving, the correlations are strongly
oscillating between the envelopes
\begin{equation}
  \label{eq:Mon16Nov175125CET2020}
  g_{\sigma,\Omega\rightarrow\infty}^{(2)}(\tau) =
  1\pm e^{-3\gamma_\sigma \tau/4}\,,
\end{equation}
that decay from~0 and~2 respectively to~1. Note also that the
expressions in Eqs.~(\ref{eq:Thu23Apr2020225637BST}),
(\ref{eq:Thu23Apr2020230012BST}--\ref{eq:Mon16Nov175125CET2020}) are
auto-correlations, and as such they are symmetric functions,
namely~$g_\sigma^{(2)}(-\tau) = g^{(2)}_\sigma(\tau)$.

Already, we have more than enough material to study from this basic
emitter the highly-nontrivial physics of loss of antibunching, and we
will focus the rest of our discussion on this case. Other antibunched
sources would either behave similarly and/or could be studied
following a similar approach. 

\section{Loss of antibunching by noise contamination}
\label{sec:Fri30Jul190600CEST2021}

A first obvious and simple way that antibunching can be lost is due to
the signal being perturbed by noise. Dark counts, for instance, which
correspond to photon-detection even in absence of light (whence the
name)~\cite{stucki01a}, clearly spoil antibunching, since the extra
photon can arrive simultaneously with a signal photon that was
supposed to be detected in isolation.  Also, in some cases, the laser
driving the system can directly inject a spurious fraction of photons
into the detectors~\cite{yuan09a}. All these photons that are
uncorrelated with the source cause a random noise, or shot noise,
which (usually) spoils antibunching (shoit noise usually assume
Poissonian statistics of the noise).

If we call $I(t)$ the instantaneous photon intensity from the source
(signal) and $I'(t)$ that of the randomly added photons (noise), the
total final intensity is given by:
\begin{equation}
  \label{I^*}
  I^*(t) = I'(t) + I(t)\,,
\end{equation}
and the photon statistics of the total signal is given by
\begin{equation}
  \label{g2random}
g^{*(2)}(\tau) = \frac{\langle{:}I^*(t)I^*(t+\tau){:}\rangle}{\langle I^*(t) \rangle ^2}\,,
\end{equation}
where ${:}\cdots{:}$ normal-orders the operators so that they appear
in the form of Eq.~(\ref{eq:Sat31Jul141755CEST2021}). Since signal and
noise are uncorrelated
$\langle{:}I(t)I'(t'){:}\rangle = \langle I(t) \rangle \langle
I'(t')\rangle$ for all~$t$, $t'$
and also introducing $\xi$ the noise to signal ratio, i.e.,
$\xi\equiv \langle I'(t)\rangle/\langle I(t) \rangle$, one can get a
simple expression that relates the photon statistics of the signal
contaminated by the noise with
statistics~$g'^{(2)}(\tau)\equiv\langle{:}
I'(t)I'(t+\tau){:}\rangle/{\langle I'(t) \rangle ^2}$ to that of the
original signal~$g^{(2)}(\tau)$~as
\begin{equation}
  \label{eq:Sat31Jul124750CEST2021}
  g^{*(2)}(\tau) = \frac{1}{(1+\xi)^2}\left( g^{(2)}(\tau) + \xi^2g'^{(2)}(\tau)+ 2 \xi \right)\,.  
\end{equation}
%
%
If the noise has no correlation, $g'^{(2)}(\tau)=1$ for all~$\tau$ and
for perfect antibunching with~$g^{(2)}(0)=0$, the loss of antibunching
$\xi(2+\xi)/(1+\xi)^2$ requires a noise-to-signal ratio of
$\sqrt{2}-1\approx 42\%$ to spoil it to 0.5 and even when there is
twice as much noise as perfectly antibunched signal, the resulting
antibunching of $8/9\approx0.89$ is still comfortably observed. The
random noise tends to flatten the correlations to that of an
uncorrelated (coherent) signal with~$g^{*(2)}(\tau)=1$ everywhere, in
a way that shifts the curves to one, without transforming bunching
into antibunching or vice versa~\cite{teich82a}. It also has no effect
on the coherence time (the time necessary for the correlation to
converge to one is independent of the percentage of noise). Thermal
noise, not surprisingly, is more detrimental to antibunching, with
$\xi=1/3$ to spoil perfect antibunching to 0.5 and when noise and
signal are equal in intensity, then $g^{*(2)}(0)=1$ with
super-Poissonian statistics for higher~$\xi$. Depending on the
coherence time of the thermal light, $g^{*(2)}(\tau)$ is either
bunched or antibunched in the sense of decreasing or increasing
correlations in time. If the noise itself is antibunched, it cannot
increase $g^{*(2)}(0)$ beyond~$0.5$, which it does
when~$\xi=1$. Further noise reduces $g^{*(2)}(0)$ again as the
original signal becomes the noise for the now dominating antibunching.
In all cases, in the limit $\xi\to\infty$,
$g^{*(2)}(\tau)\to g'^{(2)}(\tau)$ and then one observes the noise
itself, so what is lost is indeed the antibunching of the signal (such
as its coherence time). Finally, we commented already how a possible
source of noise is from the driving laser itself. We have assumed in
this discussion that noise and signal are independent and do not
interfere, so that their intensity simply add in a way reminiscent of
a classical pictures of photons as particles which are superimposed
the ones onto the others. In this respect the normal ordering above
plays no direct role and one could understand the result with
classical stochastic fields~$I$ which are not number-operators. It
could also be the case that the admixing of the two quantum fields is
done at the level of their amplitudes, in which case the description
would need to be along the lines of
Ref.~\cite{zubizarretacasalengua20b} where time and operator orderings
would be significant and more complex correlations could be obtained
as a result.

\section{Loss of antibunching by time uncertainty}
\label{sec:Fri30Jul190612CEST2021}

We now turn to the loss of antibunching due to a time uncertainty in
the detection of the photons. Such a scenario can be due to a
dead-time of the detector or a jitter effect. In both cases, the
result is that a photon arriving to the detector at a time~$t_0$ is
reported by the latter at some other time~$t_0+t$, with~$t$ following
a probability distribution~$D^2_\Gamma(t)$, to which we will refer as
the ``jitter'' function. The parameter~$\Gamma$ is the inverse of the
mean jitter time, and it is also a measure of the width of the
distribution. Independently of~$\Gamma$, since we assume perfect
detection, the jitter function must be such that
\begin{equation}
  \label{eq:Thu23Apr2020212655BST}
  \int_{-\infty}^\infty D_\Gamma^2(t)\,dt=1\,,
\end{equation}
implying that all the photons that arrive to the detector are
ultimately reported.  The temporal structure of the photon stream that
is received by the detector is modified by the jitter function. Its
effect can be formally taken into account through the physical
spectrum of emission, defined as~\cite{eberly77a}
\begin{multline}
  \label{eq:Thu23Apr2020213652BST}
  S_{\Gamma}^{(1)}(\omega,T)=\frac{1}{2\pi}\iint_{-\infty}^{\infty}
  D_{\Gamma}(T-t_1) D_{\Gamma}(T-t_2)\\
  e^{i\omega(t_2-t_1)}\langle\ud{a}(t_1)a(t_2)\rangle\,dt_1dt_2\,,
\end{multline}
where~$a$ is the annihilation operator of the field emitting the
photons and~$\omega$ is the frequency at which these photons are being
emitted. Integrating Eq.~(\ref{eq:Thu23Apr2020213652BST}) in frequency
leads to the time-resolved population
\begin{equation}
  \label{eq:Thu23Apr2020213503BST}
  S_{\Gamma}^{(1)}(T)=  \int_{-\infty}^{\infty}
  D^2_{\Gamma}(T-t)\langle\ud{a} a\rangle (t)\,dt \,.
\end{equation}
Namely, the time-resolved population of the light emitted is the
convolution of the population with the jitter function. If the signal
is in a steady state, then the population is independent of time,
$\mean{\ud{a}a}=n_a$, and we find that the time-resolved population is
equal to the total intensity, $S_\Gamma(T) = n_a$, and that the jitter
does not play a role on this observation.  This makes sense since in
the steady state, shuffling the times at which the photons are reported
does not change the mean number of detected photons per unit time.

Applying the same treatment to the intensity-intensity correlation
function describes the effect of the time jitter on~$g^{(2)}(\tau)$
which now persists even in the steady state.  Instead of the physical
spectrum, one starts in this case with the second-order correlation
function resolved in time and frequency~\cite{delvalle12a}:
\begin{multline}
  \label{eq:Thu23Apr2020214433BST}
  S_{\Gamma_1\Gamma_2}^{(2)}(\omega_1,T_1;\omega_2,T_2)=\frac{
    \Gamma_1\Gamma_2}{(2\pi)^2}\iiiint_{-\infty}^{\infty}dt_1  
  dt_2 dt_3  dt_4\\ D_{\Gamma_1}(T_1-t_1)  D_{\Gamma_1}(T_1-t_4)
  D_{\Gamma_2}(T_2-t_2)   D_{\Gamma_2}(T_2-t_3) \\ 
  e^{i\omega_1(t_4-t_1)}e^{i\omega_2(t_3-t_2)} \langle
  \mathcal{T}_-[\ud{a}(t_1)\ud{a}(t_2)]\mathcal{T}_+[a(t_3)a(t_4)]\rangle 
  \,.
\end{multline}
where~$\Gamma_1$ and~$\Gamma_2$ accommodate the fact that the pair of
photons can be recorded by detectors with different jitters, and where
$\mathcal{T}_\pm$ means time-reordering of the operators so that the
first time is to the far right/left, and the whole expression is
time-ordered. Assuming that the uncertainty for both detectors is
equal (i.e.,~$\Gamma_1=\Gamma_2=\Gamma$), the frequency-integrated
quantity becomes:
\begin{multline}
  \label{eq:Thu23Apr2020220154BST}
  S_{\Gamma}^{(2)}(T_1;T_2)=
  \iint_{-\infty}^{\infty} D^2_{\Gamma}(T_1-t_1) D^2_{\Gamma}(T_2-t_2) \\
  \langle\mathcal{T}_-
  [\ud{a}(t_1)\ud{a}(t_2)]\mathcal{T}_+[a(t_2)a(t_1)]\rangle\,dt_1 
  dt_2\,, 
\end{multline}
which corresponds to the intensity correlations between two branches
of a split-signal that is affected by the time jitter. Time ordering
of the operators leads to two integration regions, $t_1>t_2$
and~$t_2>t_2$, that allows to express
Eq.~(\ref{eq:Thu23Apr2020220154BST}) in terms of the
standard~$G^{(2)}(t,x)$ of the system,
letting~$t=\mathrm{min}(t_1,t_2)$ and~$x=|t_1-t_2|$:
\begin{multline}
  \label{eq:Thu23Apr2020222540BST}
  S_{\Gamma}^{(2)}(T_1;T_2)= \\
  \int_{-\infty}^{\infty}dt\int_0^\infty dx\,
  G^{(2)}(t,x)
  \left[ D^2_{\Gamma}(T_1-t )
  D^2_{\Gamma}(T_2-t-x)\right.  +{}\\ \left. {}+ D^2_{\Gamma}(T_1-t-x)
  D^2_{\Gamma}(T_2-t)\right]\,,
\end{multline}
where~$0\leq x <\infty$ represents the delay between the two detected
photons.  The second-order correlations function of the stream of
photons affected by the time jitter is then obtained by normalizing
Eq.~(\ref{eq:Thu23Apr2020222540BST}) with
Eq.~(\ref{eq:Thu23Apr2020213503BST}) as
\begin{equation}
  \label{eq:Tue20Jun170402BST2017}
  g_{\Gamma}^{(2)}(T_1;T_2)=S_{\Gamma}^{(2)}(T_1;T_2)/[S_{\Gamma}^{(1)}(T_1)
  S_{\Gamma}^{(1)}(T_2)]\,.   
\end{equation}
In the case of a steady state, it is convenient to express the
correlations as a function of the time-delay between the photon pairs
(the initial time is irrelevant in this case). This provides the main
result of this Section in the form of the jittered~$g_\Gamma^{(2)}$
expression from that of the original signal as affected by the jitter
function~$D_\Gamma$:
%
%
\begin{multline}
  \label{eq:Fri24Apr2020113650BST}
  g_{\Gamma}^{(2)}(\tau)=\\
  \int_0^\infty g^{(2)}(\theta)\int_{-\infty}^{\infty} \left[
    D^2_{\Gamma}(-t-\theta) D^2_{\Gamma}(\tau-t) \right. +{}\\ \left.
    {} +    D^2_{\Gamma}(-t ) D^2_{\Gamma}(\tau-t-\theta)\,dt\right]\,d\theta\,,
\end{multline}
where we have used the fact that in the steady
state~$S_\Gamma^{(1)}(T)=n_a$ is independent of time,
and~$g^{(2)}(\theta)$ is the second-order correlation of the photon
stream \emph{without} the time-jitter.
%
%
Equation~(\ref{eq:Fri24Apr2020113650BST}) is general: it holds for
any~$g^{(2)}(\theta)$ (as long as it is obtained in the steady state), and
the particular shape of the correlations with time jitter will be
given by the rightmost integral, which depends only on the jitter
function.

We now consider a few particular cases for the jitter function,
describing a possible physical origin in each case.  The analytical
expressions for the corresponding photon correlations for the emission
from a two-level system (2LS) in the various regimes of
excitation---which generalize the fundamental
expressions~(\ref{eq:Thu23Apr2020225637BST})~\&~(\ref{eq:Thu23Apr2020230012BST})---can
be obtained in all theses cases, but they are bulky and not
enlightening per se, therefore we provide their full expression in the
appendix. All the distributions have been chosen such that their
variance be identical, namely, equal to~$1/\Gamma^2$, so as to compare
them usefully (the variance is a better indicator of the effect of the
distribution on antibunching than the mean).

\begin{figure}
  \includegraphics[width=.85\linewidth]{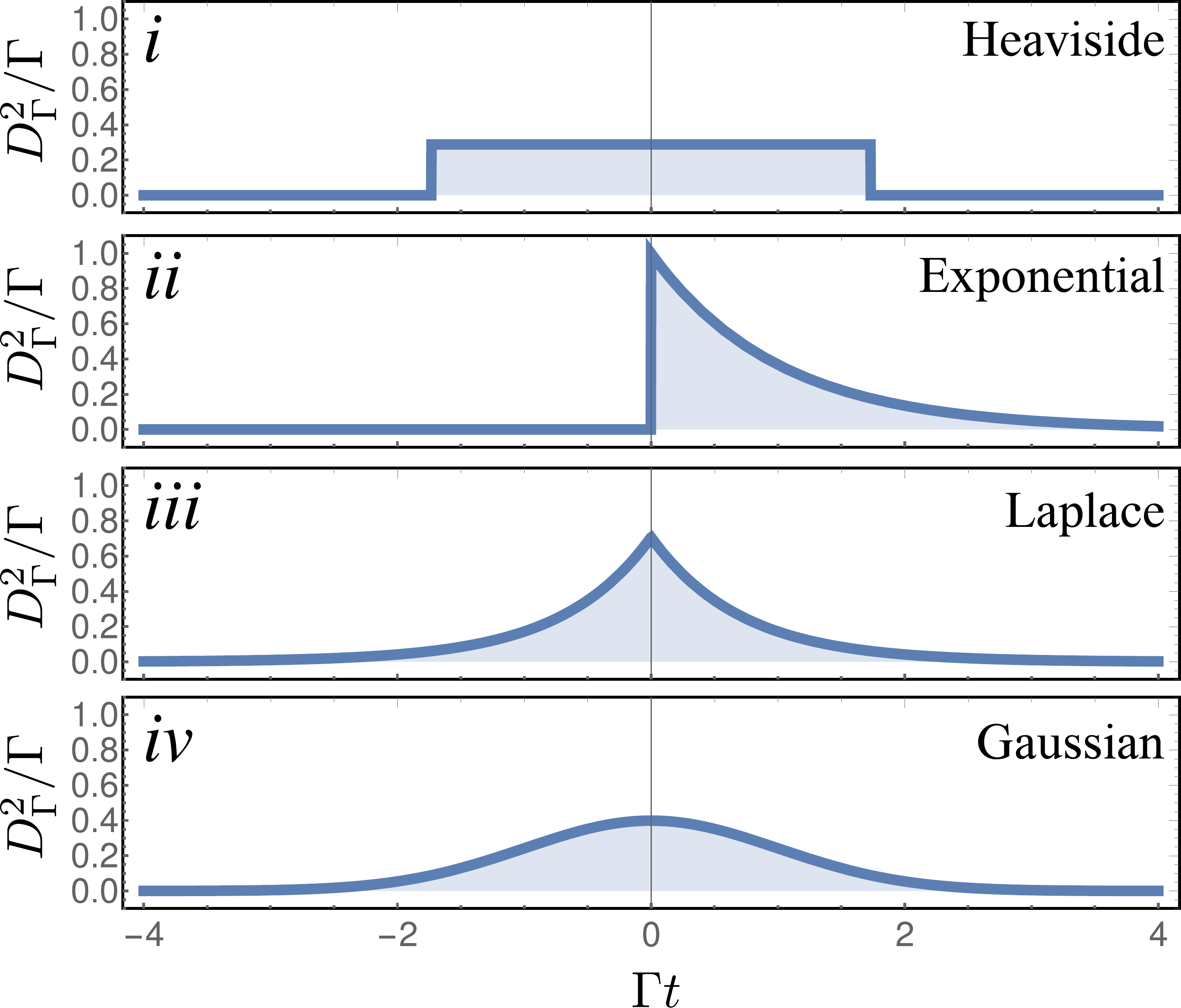} 
  \caption{Jitter functions considered in this text, namely
    ($i$)~Heaviside, Eq.~(\ref{eq:Fri24Apr2020111614BST}),
    ($ii$)~Single exponential, Eq.~(\ref{eq:Tue28Apr2020130804BST}),
    ($iii$)~Double exponential (Laplace),
    Eq.~(\ref{eq:Fri24Apr2020231635BST}) and ($iv$)~Gaussian,
    Eq.~(\ref{eq:Sat25Apr2020195622BST}), all with the
    variance~$1/\Gamma^2$. A photon from the original signal at
    time~$0$ is replaced by one at time~$t$ following the respective
    distributions.}
    \label{fig:Mon4May2020130705BST}
\end{figure}

\begin{figure*}[t]
  \includegraphics[width=\linewidth]{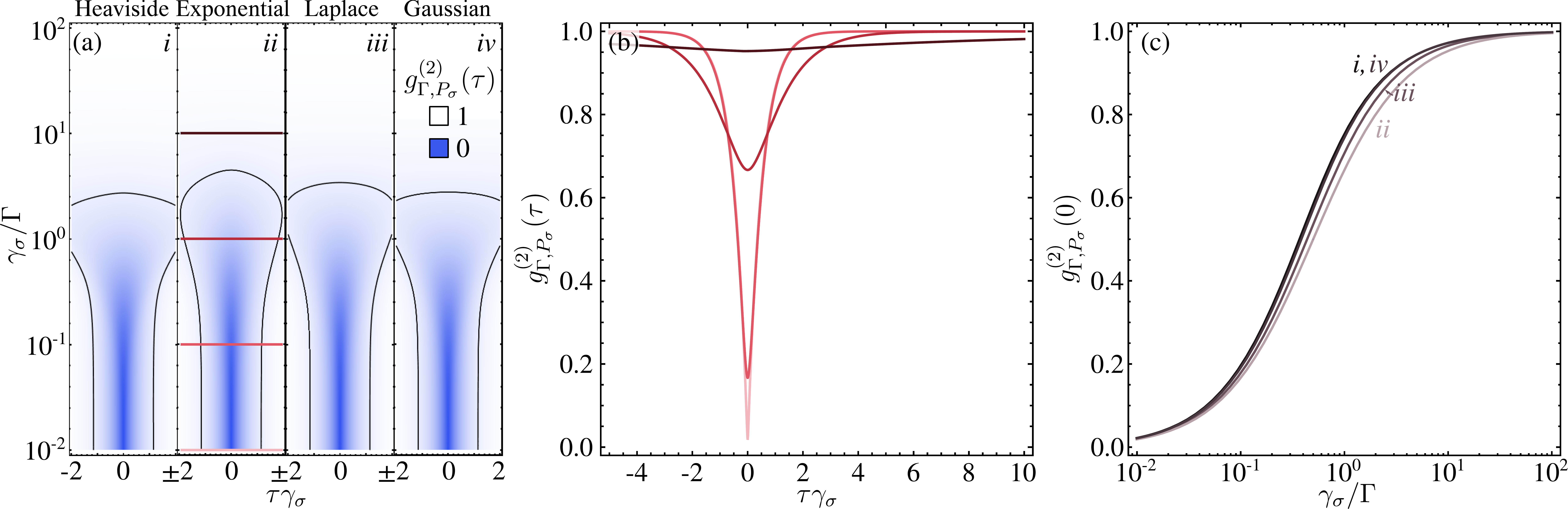}
  \caption{Photon-correlations with various types of time jitter for
    the incoherently driven 2LS. (a)~Transition from perfect
    antibunching when the mean jitter time~$\tau_j$ vanishes
    (corresponding to a large~$\Gamma$ of the jitter functions) to
    uncorrelation. The progression occurs at different speeds
    depending on the type of jitter, namely, $i$~Heaviside,
    $ii$~Double exponential, $iii$~Single exponential
    and~$iv$~Gaussian. The black contours shows the
    isolines~$g^{(2)}_{\Gamma,P_\sigma}=0.9$ (b)~Cuts along the lines
    marked in~(a)$iii$ (single-exponential). (c)~Zero delay
    correlations for the various types of jitter. All figures are
    for~$P_\sigma=\gamma_\sigma$.}
    \label{fig:Mon4May2020180030BST}
\end{figure*}
\begin{figure*}[t]
  \includegraphics[width=\linewidth]{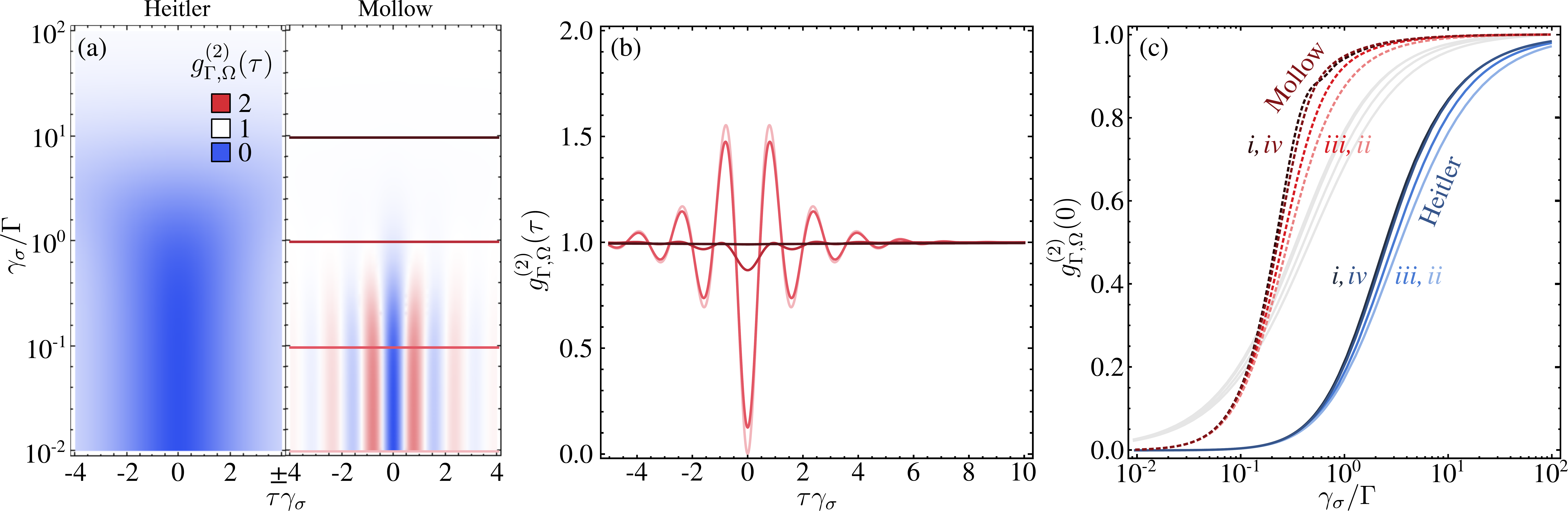}
  \caption{Photon-correlations with various types of time jitter for
    the coherently driven 2LS. (a)~The loss of antibunching depends
    strongly on the regime of excitation, i.e., in the Heitler regime
    (left) the transition to uncorrelation requires a greater time
    jitter than in the Mollow triplet regime (right). The behavior
    with the four jitter functions follows the trend shown in
    Fig.~\ref{fig:Mon4May2020180030BST}(a), and therefore we only show
    the case for the single exponential. (b)~Cuts along the lines
    marked on (a) for the Mollow triplet, showing the dampening of the
    correlations as the mean jitter time increases. (c)~Zero delay
    correlations: the solid and dashed lines correspond to the 2LS
    driven in the Mollow and Heitler regimes, respectively. The
    various lines~$i$--$iv$ correspond to the different jitter
    functions, namely, Heaviside, double exponential, single
    exponential and Gaussian, respectively. Figures are
    for~$\Omega=\textcolor{red}{??}$ in the Heitler regime and
    $\Omega=2\gamma_\sigma$ in the Mollow regime.}
  \label{fig:Tue17Nov112503CET2020}
\end{figure*}
%

%

\begin{enumerate}
\item The \textbf{Heaviside function} describes a device that has no
  time resolution within a given time window, in which case, the jitter
  function reads
\begin{equation}
  \label{eq:Fri24Apr2020111614BST}
  \kern.75cm D_\Gamma(t)\equiv\sqrt{\Gamma\over\sqrt{12}}\theta\left(1 -{2\Gamma t\over\sqrt{12}}\right)\theta\left({1} +{2\Gamma t\over{\sqrt{12}}}\right)\,,
\end{equation}
where~$\theta(t)$ is the Heaviside function, and the distribution in
Eq.~(\ref{eq:Fri24Apr2020111614BST}) is only nonzero in the
interval~$|t|<\sqrt{3}/\Gamma$ (chosen, again, so that the variance of
the jitter is $1/\Gamma^2$).  This could correspond to a streak camera
which randomizes the time information within one pixel of the CCD
camera~\cite{silva16a}. The filtered correlations are given by
Eq.~(\ref{eq:Tue17Nov093639CET2020}) for incoherent excitation and by
Eq.~(\ref{eq:Tue28Apr2020213841BST}) for coherent excitation.
\item The \textbf{Exponential function} describes a device that is
  equally likely to trigger the signal at any moment that follows its
  excitation, with jitter function
\begin{equation}
  \label{eq:Tue28Apr2020130804BST}
  D_\Gamma(t)=\sqrt{\Gamma} \theta(t)e^{-\Gamma t/2}\,,
\end{equation}
where~$\theta(t)$ is the Heaviside function, as shown in
Fig.~\ref{fig:Mon4May2020130705BST}(b). This describes a wide class of
devices with a memoryless dead time. The filtered correlations are
given by Eq.~(\ref{eq:Tue28Apr2020132411BST}) for incoherent
excitation and by Eq.~(\ref{eq:Tue28Apr2020135027BST}) for coherent
excitation.
\item The \textbf{double exponential function}, also known as Laplace
  distribution, describes a device that has a memoryless dead time not
  only in its signal emission, like the previous type, but also in its
  excitation time, with jitter function
  \begin{equation}
    \label{eq:Fri24Apr2020231635BST}
    D_\Gamma(t)=\sqrt{{\Gamma\over\sqrt{2}}} e^{-|t|\Gamma/\sqrt{2}}\,,
  \end{equation}
  shown in Fig.~\ref{fig:Mon4May2020130705BST}(c). It can be seen as a
  refinement of the two previous cases, where the variation at which
  the photons are reported can be both delayed or advanced according
  to an exponential function. The filtered correlations are given by
  Eq.~(\ref{eq:Tue17Nov110858CET2020}) for incoherent excitation and
  by Eq.~(\ref{eq:Sat25Apr2020192637BST}) for coherent excitation.
\item Finally, the \textbf{Gaussian function} describes
  normally-distributed fluctuations in the detection time, with jitter
  function
\begin{equation}
  \label{eq:Sat25Apr2020195622BST}
  D_\Gamma(t)=\frac{\sqrt{\Gamma}}{(2\pi)^{1/4}} e^{-(\Gamma t/2)^2}\,.
\end{equation}
This could be due to, e.g., the electronics involved in the detection
of a photon after its arrival, or various types of
noise~\cite{ulrich07a}. The filtered correlations are given by
Eq.~(\ref{eq:Sat25Apr2020201941BST}) for incoherent excitation and by
Eq.~(\ref{eq:Mon27Apr2020205804BST}) for coherent excitation.
\end{enumerate}

%

The two-photon correlations as seen by a detector with the four types
of time jitters just described are shown in
Fig.~\ref{fig:Mon4May2020180030BST} for incoherent excitation and in
Fig.~\ref{fig:Tue17Nov112503CET2020} for coherent excitation, as
follows from the analytical expressions given in the appendix.

The most striking result is that while various types of jitter result
in an overall identical loss of antibunching, the robustness of the
photon correlations depends on the type and the regime of driving,
even though the emitter is the same (2LS), namely, the most robust
photon correlations are from the coherent driving in the Heitler
regime. This can be understood to some extent from the coherence time
of these correlations, which is shorter than for the incoherently
driven 2LS. Comparing the panels~(c) in
Figs.~\ref{fig:Mon4May2020180030BST}
and~\ref{fig:Tue17Nov112503CET2020} (the trace of the former are
reproduced in light-gray in the latter), one can see how the Mollow
regime, the incoherent 2LS and finally the Heitler regime appear in
order of least robust to most robust (the same amount of any type of
jitter affects much more the Mollow antibunching than it does its
Heitler counterpart). For very small filtering,
with~$\Gamma/\gamma_\sigma\lessapprox 0.1$, the Mollow antibunching
gets more robust than the incoherent 2LS, which is very fragile to
imperfect detection. It is well-known that resonant excitation yields
stronger antibunching but this is attributed to the cleaner
environment that is free of carriers, heating, etc. Here we find that
at a fundamental level too, resonant excitation typically produces a
stronger antibunching in the sense that it is more resilient to
factors that spoils it.  The Heaviside and Gaussian types of jitter
lead to almost identical losses of antibunching, suggesting that
spreading more photons has a worse effect than displacing fewer of
them but farther. There is, nevertheless, a difference between the
Haviside and Gaussian types of jitter in the Mollow case, since the
former filters resolves the side peaks ``all of a sudden'', while the
Gaussian perceives them in advance, showing as well the importance of
the spectral structure. This produces, when that happens, a ripple in
the loss of antibunching with heaviside-type of jitter. In all cases,
the single-exponential, memory-less jitter, is the one that least
affects the antibunching. Since the exact, quantitative results differ
only slightly from one type of jitter to the other, we show the traces
for the loss of antibunching for one case only ($ii$,
single-exponential) in Panels~(b), where the temporal dependence also
exhibits a change in the correlation time in addition to the damping
of the correlations.  In all cases, in the limit of large widths of
the jitter noise, the correlations follow a Poisson distribution. The
mean uncertainty brought to the times at which the photons are
observed increases to a point where times are essentially
randomized. While this is true regardless of the type of jitter, the
speed of this randomization varies.

As a concluding remark, while there is, obviously, a link between the
type of time uncertainty and the observed correlations, this does not
follow from a mere convolution of the naked signal with the noise and
involves some specifics of the dynamics (such as the regime of
driving). Therefore, the original correlations cannot be recovered
simply by deconvolution of the raw data, and while we do not discuss
here how serious or superficial is this problem, or how it could be
best compensated for, care should be exerted regarding the methods
reported in the literature that seem to overlook these
aspects~\cite{michler00a, fleury00a,kurtsiefer00a,messin01a, flagg09a,
  nothaft12a, konthasinghe12a, matthiesen12a, he13a, reithmaier15a,
  he15a, kumar16a, malein16a, gao17a, snijders18a, zhao18a, fink18a,
  liu18a, 
  foster19a, liu19a, anderson20a, phillips20a}.


\section{Loss of antibunching by frequency filtering}
\label{sec:Fri30Jul190644CEST2021}

Antibunching is also spoiled by another reason, this time more
fundamental since it is inherent to any detection process. Consider a
photon counting experiment in which one not only has the information
about the time of arrival, but also about the frequency (or energy) of
the detected photon. In such a scenario, Heisenberg's uncertainty
principle applies: the absolute uncertainty about the frequency of the
photons allows to have a perfect resolution in the time of arrival
(and therefore in the time of emission) of each photon. Satisfying
this condition allows to observe perfect antibunching, namely, for a
2LS, the second-order correlation functions given in
Eq.~(\ref{eq:Thu23Apr2020225637BST}) and
Eq.~(\ref{eq:Thu23Apr2020230012BST}) for incoherent and coherent
driving, respectively. However, gaining information about the
frequency of the detected photons inevitably means that the temporal
resolution ceases to be perfect, and antibunching gets lost as a
consequence. Every detectors have a finite bandwidth and emission
spectra are typically Lorentzian, i.e., with fat tails and regularly
emission arbitrarily far from the central frequency. Therefore, all
detection is limited in its frequency range, or, equivalently,
temporal resolution. Formally, such an effect is embodied in the
expressions for the first- and second-order physical spectra, shown in
Eq.~(\ref{eq:Thu23Apr2020213652BST}) and
Eq.~(\ref{eq:Thu23Apr2020214433BST}), respectively. Thus, the
frequency-resolved second-order correlation function is then obtained
as
\begin{equation}
  \label{eq:Wed6May2020204731BST}
  g^{(2)}_{\Gamma_1\Gamma_2}(\omega_1,T_1; \omega_2,T_2) =
  \frac{S^{(2)}_{\Gamma_1 \Gamma_2}(\omega_1,T_1;
    \omega_2,T_2)}{S^{(1)}_{\Gamma_1 }(\omega_1,T_1) S^{(1)}_{\Gamma_2
    }(\omega_2,T_2)}\,,
\end{equation}
which is the counterpart of Eq.~(\ref{eq:Tue6Jun133237BST2017})
for~$n=2$ and corresponding to photons that have been filtered in
frequency before their collection.

Obtaining the frequency-resolved correlations in
Eq.~(\ref{eq:Wed6May2020204731BST}) is a complicated task, for which
several integrals must be performed, keeping track of the integration
regions that stem from the time-ordering requirements. Even for the
most fundamental system, the two-level system, and a detection
function with an exponential profile (as given in
Eq.~(\ref{eq:Tue28Apr2020130804BST})), this endeavor required some
approximations that, however, allowed to obtain approximate analytical
expressions~\cite{cohentannoudji79a, reynaud83a, dalibard83a,
  arnoldus84a, arnoldus86a, nienhuis93a, joosten00a}.

A theory to compute $N$-photon frequency-resolved photon
correlations~\cite{delvalle12a}---for which
Eq.~(\ref{eq:Wed6May2020204731BST}) is a particular case with~$N=2$---
was shown to be both exact and simple to implement. It simply consists
in adding detectors to the source of light, whose correlations model
those of the filtered emission of the source. This is exact provided
that the dynamics of the detectors does not perturb that of the source
itself. This can be ensured either by having the source and the
detector coupled with a vanishing strength (this was actually the
approach taken in the original paper~\cite{delvalle12a}) or by turning
to the cascaded formalism~\cite{gardiner93a,carmichael93a}, by which
the coupling is unidirectional and the detector becomes the target of
the excitation of the source of light, which remains unaffected by the
presence of the detector. We have also shown that, for Glauber
correlations, these two approaches are
equivalent~\cite{lopezcarreno18a} (to measure, say, the numerators of
the Glauber correlators, of which the population is a particular case
of interest, the cascading formalism must be used).  In practical
terms, the description of the augmented systems, consisting of the
source and the detector of the emission, is done through the same
master equation as~Eq.~(\ref{eq:Mon11May2020131426BST}). Thus, if the
source of light is described by a field with annihilation
operator~$\sigma$ (which in our case is the two-level system but could
be any other operator with any given quantum algebra), and the
detector is described with a bosonic operator~$\xi$, then the
Hamiltonian in Eq.~(\ref{eq:Mon11May2020131426BST}) is given by (still
with~$\hbar=1$)
\begin{equation}
  \label{eq:Thu7May2020210751BST}
  H = H + \omega_\xi \ud{\xi}\xi + \epsilon(\ud{\sigma}\xi + \ud{\xi}\sigma)\,,
\end{equation}
where~$H$ is the free Hamiltonian of the source, $\omega_\xi$ is the
frequency at which the detector is collecting the light,
and~$\epsilon$ is the (vanishing) strength of the coupling between the
source and the detector. In addition, the bandwidth~$\Gamma$ of the
detector~$\xi$ is included thought a Lindblad terms
$\mathcal{L}_{\xi} \rho = (\Gamma/2)(2 \xi \rho \ud{\xi} - \ud{\xi}
\xi \rho - \rho \ud{\xi} \xi)$. Given that the operator~$\xi$
describes the detector of the light, the parameter~$\Gamma$ can also
be interpreted as the \emph{linewidth} of the detector. Note that if
the detector were in isolation (without coupling it to the source of
light), its emission spectrum is given by a Lorentzian of
width~$\Gamma$ centered at~$\omega_\xi$. Therefore the frequency
filtering done with this detector has a Lorentzian profile, and the
time-resolution is lost following an exponential distribution with
mean~$1/\Gamma$, as the function that we have considered in
\S\ref{sec:Thu23Apr2020212320BST}. the method outlined above can also
be used to describe detectors with different temporal and spectral
resolutions, such as the ones that we have considered in the previous
section: one simply has to couple the source of light to a quantum
object, whose emission spectrum has the desired shape. However, their
implementation is more involved~\cite{kamide15a} and therefore in this
paper we will restrain to the case of Lorentzian filters.

The rest of the section is devoted to the explicit application of the
theory of frequency-resolved correlations to the emission of a
two-level system, considering both the excitation from an incoherent
and a coherent source of light.

\begin{figure*}[t]
  \includegraphics[width=\linewidth]{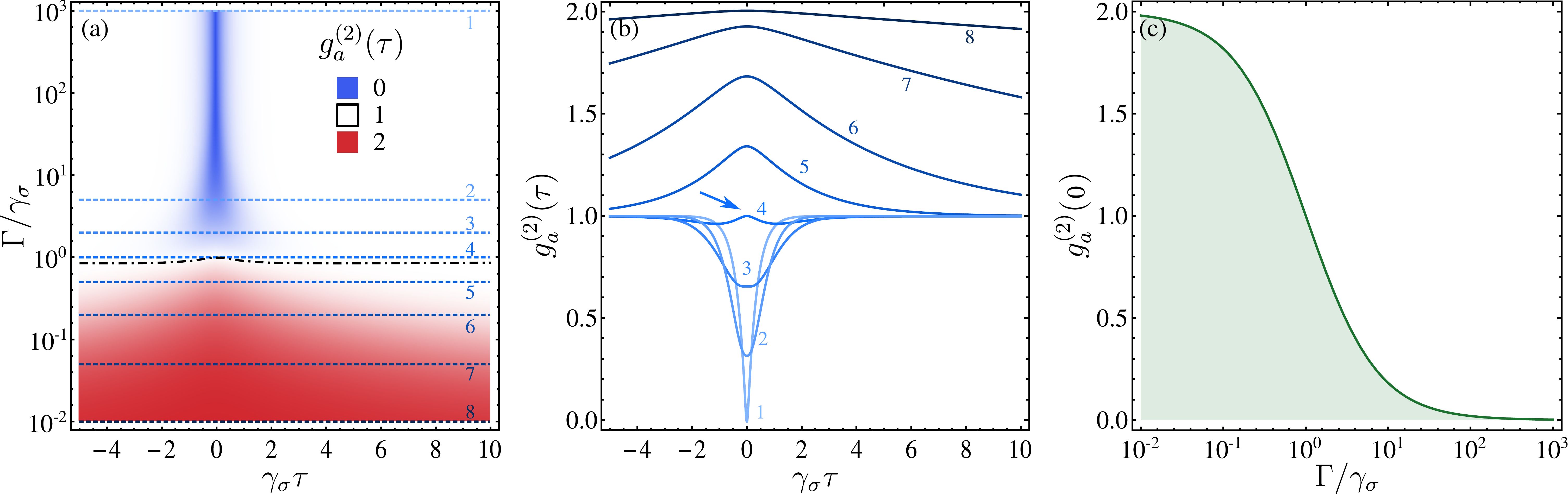} 
  \caption{Loss of antibunching of an incoherently driven 2LS due to
    frequency-filtering. (a)~The observed correlations transit from
    perfect antibunching in the limit in which the frequency of the
    detected photons is not know ($\Gamma \rightarrow \infty$) to a
    complete thermalization in the opposite
    regime~($\Gamma\rightarrow 0$), as predicted by
    Eq.~(\ref{eq:Fri28Apr215857BST2017}). (b)~Cuts of the map in
    Panel~(a) for the~$\Gamma/\gamma_\sigma$ ratios highlighted by the
    dashed blue lines shown with small colored numbers 1--8. Note that
    while the correlations vary from antibunching to bunching, they
    don't pass through an exactly uncorrelated emission, as is
    evidenced by the lump in the correlations marked by the arrow on
    line~4. (c)~Zero-delay correlation as a function of the linewidth
    of the detector, showing the smooth loss of antibunching due to
    frequency filtering.}
    \label{fig:Fri8May2020215520BST}
\end{figure*}

\subsection{incoherent excitation}

The master equation in Eq.~(\ref{eq:Mon11May2020131426BST}) with the
Hamiltonian in Eq.~(\ref{eq:Fri8May2020195906BST}) complemented with a
detector~$\xi$ of bandwidth~$\Gamma$ gives access to the frequency-resolved
correlations of the incoherently-driven 2LS. When the detector is in
resonance to the 2LS, i.e.,~$\omega_\xi = \omega_\sigma$, the
correlations become
\begin{widetext}
  \begin{equation}
    \label{eq:Fri28Apr215857BST2017}
    g_{\sigma}^{(2)}(\tau) = 1 - \left(
      \frac{\Gamma}{\Gamma-\Gamma_\sigma} \right)^2 e^{-\Gamma_\sigma
      \tau} + \frac{\Gamma_\sigma(\Gamma_\sigma^2-3\Gamma
      \Gamma_\sigma-2\Gamma^2)}{(\Gamma_\sigma-\Gamma)^2
      (\Gamma_\sigma+3\Gamma)} e^{-\Gamma \tau}+\frac{2\Gamma_\sigma
      \Gamma (5\Gamma-\Gamma_\sigma)}{(\Gamma_\sigma-\Gamma)^2
      (\Gamma_\sigma+3\Gamma)} e^{-(\Gamma_\sigma+\Gamma)\tau/2}\,,
  \end{equation}
\end{widetext}
where $\Gamma_\sigma = \gamma_\sigma + P_\sigma$. This result has been
confirmed by independent numerical calculations from
frequency-resolved Monte Carlo simulations in
Ref.~\cite{lopezcarreno18a}. The expressions for higher-order
correlations can also be obtained in a closed form, but they become
bulky and typically not considered for antibunching so we do not
consider them here. Furthemore, in Ref.~\cite{lopezcarreno16b}, we
have provided a recurrence relation for all correlations at zero
delay.  The expression in Eq.~(\ref{eq:Fri28Apr215857BST2017}) is
written for positive values of~$\tau$, but these correlations are a
symmetric function of time,
namely~$g_{\sigma}^{(2)}(\tau) = g_{\sigma}^{(2)}(-\tau)$, as is shown in
Fig.~\ref{fig:Fri8May2020215520BST}. There, one can appreciate the
transition form perfect antibunching (in the limit in which the
linewidth of the detector is infinite, $\Gamma \rightarrow \infty$,
and one does not have information about the frequency of the observed
photons), to photon bunching (in the opposite regime, where the
linewidth of the detector is much smaller than the linewidth of the
emission, $\Gamma/\gamma_\sigma \rightarrow 0$) where the extreme
frequency filtering yields a thermalization of the
signal~\cite{centenoneelen92a, gonzaleztudela13a, silva16a,
  lopezcarreno16b}. Cuts for some values of the detector linewidth are
shown in Panel~(b), where the various lines (marked with numbers) have
a corresponding horizontal dashed line in panel~(a). There is a
transition of the correlations from antibunching to a thermal state as
filtering tightens with an increase of the coherence time. These
results have also been previously confirmed through a
frequency-resolved Monte Carlo numerical
experiment~\cite{lopezcarreno18a}.  Note that the
isoline~$g^{(2)}_a(\tau)=1$ is not straight (it is shown as a black
dashed-dotted line in the density plot of
Fig.~\ref{fig:Fri8May2020215520BST}(a)), and therefore, the passage
from antibunching to bunching does not transit through exactly
uncorrelated (or coherent light). The shape is similar to that of an
antibunched signal contaminated by thermal noise with a smaller
coherence time (cf.~Eq.~(\ref{eq:Sat31Jul124750CEST2021})) and
possibly the effect of the filter could be understood in these terms:
as converting some photons from the signal by thermal photons. The
particular case of Eq.~(\ref{eq:Fri28Apr215857BST2017}) at zero-delay
reduces~to
\begin{equation}
  \label{eq:Fri28Apr235011BST2017}
  g^{(2)}_\sigma(0)=\frac{2\Gamma_\sigma}{\Gamma_\sigma+3\Gamma}\,,
\end{equation}
and is shown in Fig.~\ref{fig:Fri8May2020215520BST}(c) (this was also
obtained from the cascaded formalism in Ref.~\cite{lopezcarreno16a},
Eq.~(14b)), where we can appreciate the smooth transition from
antibunching to bunching, while also recovering the universal
behaviour of thermalization by extreme filtering and of no filtering
when detecting at all frequencies, namely
\begin{equation}
  \label{eq:Sat29Apr081512BST2017}
  \lim_{\Gamma\rightarrow0}g^{(2)}_\sigma(\tau)=2
  \quad\mathrm{and}\quad
  \lim_{\Gamma\rightarrow\infty}g^{(2)}_\sigma(\tau)=g^{(2)}(\tau)\mathrm{\ of\,\,Eq.~(\ref{eq:Thu23Apr2020225637BST})}\,.
\end{equation}
A series expansion of~$\Gamma$ around zero give the time-dependence
for the correlation function for vanishing filters as
$g^{(2)}(\tau)\approx 1+\exp(-\Gamma\tau)$, showing how in this case
the dynamics of the emitter itself is completely washed out by the
filter which is sole responsible for the statistics of the surviving
photons.  The limit of vanishing frequency-filtering thus deviates
qualitatively from the case considered in the previous Section, where
the loss of antibunching was due to temporal uncertainty in the
measurement. Here, we observe thermalization,
with~$g_\sigma^{(2)}(0)=2$, while in the latter case we observed that
the signal became uncorrelated,
with~$\lim_{\Gamma \rightarrow 0} g^{(2)}_\Gamma (\tau)=1$, regardless
of the jitter function. This highlights the fundamental difference
between the two mechanisms through which the antibunching is lost,
with the randomization in the case of temporal uncertainty and the
indisitnguishablility bunching stemming from
frequency-filtering~\cite{gonzaleztudela13a}.

The transition is thus richer than one could have expected, although
the overal physical behavior matches with expectations. Therefore,
Eq.~(\ref{eq:Sat29Apr081512BST2017}) should be more accurately used
than the customary single exponential
fit~$g^{(2)}(\tau)\approx 1-g^{(2)}_0\exp(-t/\tau_0)$, where the value
of the parameter~$g_0$ is introduce from a ``deconvolution'' of the
photon correlations with the temporal profile of the instrument
response function (IRF)~\cite{michler00a,
  fleury00a,kurtsiefer00a,messin01a, flagg09a, nothaft12a,
  konthasinghe12a, matthiesen12a, he13a, reithmaier15a, he15a,
  kumar16a, malein16a, gao17a, snijders18a, zhao18a, fink18a,
  liu18a, 
  foster19a, liu19a, anderson20a, phillips20a}. The observation of
this detailed structure of the loss of antibunching seems to be
readily observable experimentally and would consisted a fundamental
test of the theory of frequency-resolved photon correlation from the
one of the most basic types of quantum emission.

\begin{figure*}[ht]
  \includegraphics[width=\linewidth]{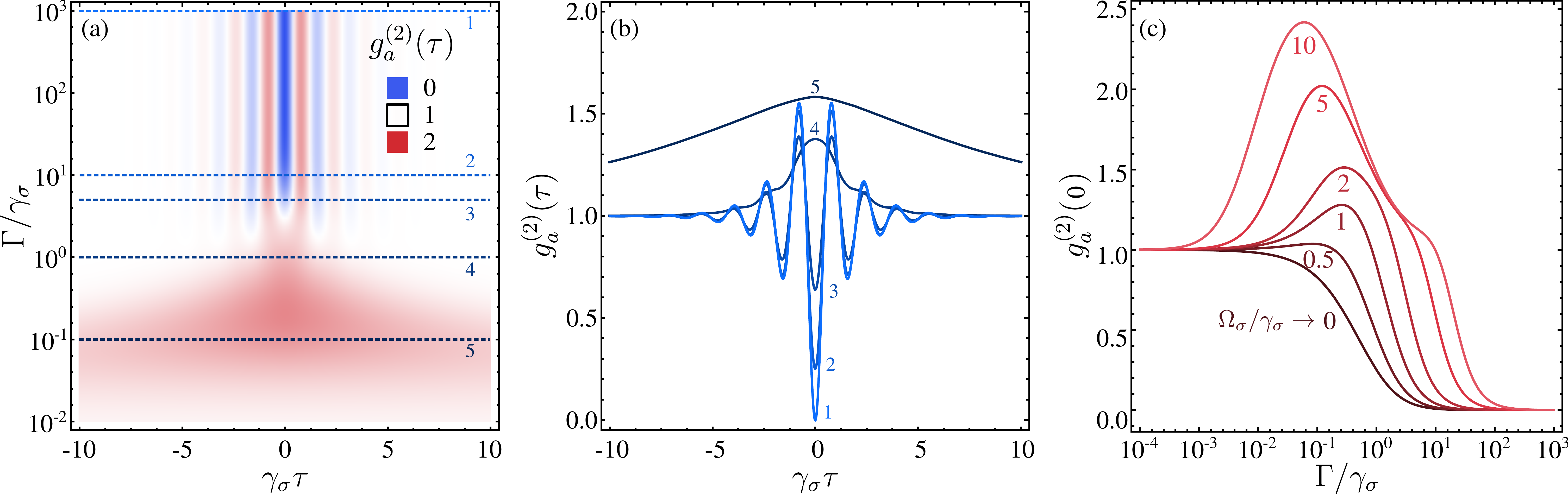} 
  \caption{Loss of antibunching of a coherently driven 2LS due to
    frequency-filtering. (a)~The Rabi oscillations induced by the
    laser are dampened by the frequency filtering, and the perfect
    antibunching obtained in the limit~$\Gamma\rightarrow \infty$
    thermalises when~$\Gamma=\gamma_\sigma$. However, further reducing
    the linewidth of the detector reveals the uncorrelated nature of
    the light emitted by the driving laser. (b)~Cuts of the density
    plot in panel~(a) illustrating the transition between
    regimes. (c)~The zero-delay correlation of the light emitted by
    the coherently driven 2LS can be tuned through the linewidth of
    the detector. The figure corresponds
    to~$\Omega_\sigma=2\gamma_\sigma$. }
  \label{fig:Mon11May2020173914BST}
\end{figure*}

\subsection{Coherent excitation}

\begin{figure*}[t]
  \includegraphics[width=.8\linewidth]{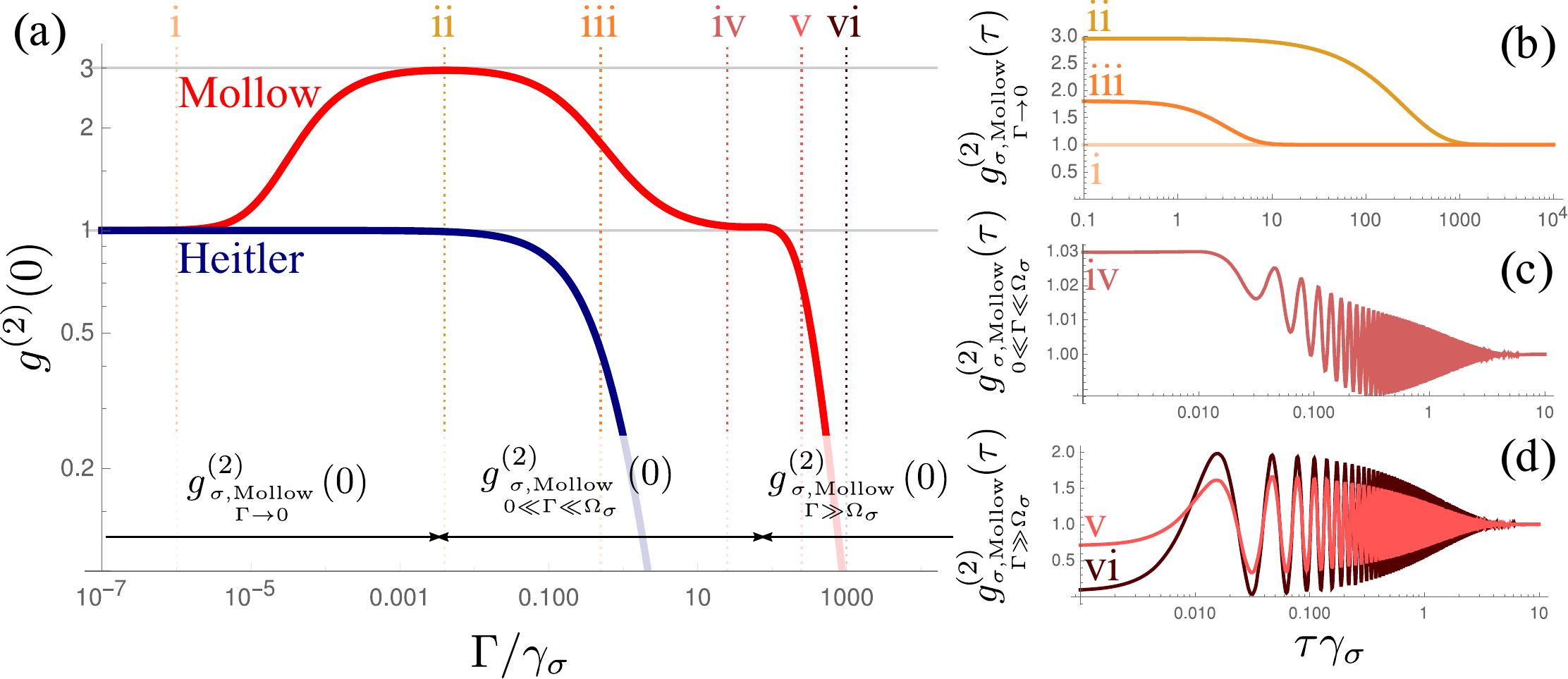}   
  \caption{Loss of antibunching for the coherently-driven 2LS both at
    (a) zero-delay and (b--d) as a function of time~$\tau$ in the
    various limits of the Mollow triplet
    ($\Omega_\sigma\gg\gamma_\sigma$ depending on the filter widths:
    (b) when $\Gamma$ is small, exhibiting bunching, (c) at
    intermediate~$\Gamma$, with onset of correlations and~(d) when
    $\Gamma$ is large and beyond the triplet splitting, exhibiting
    antibunching. Note that the case~iii which belongs to the
    intermediate regime displays a correlation function akin to those
    of the low-$\Gamma$ case and is best represented with them. These
    results can be obtained either from
    Eqs.~(\ref{eq:Wed3Jun2020232311BST})--(\ref{eq:Fri6Aug173936CEST2021})
    in the corresponding limits or from
    Eq.~(\ref{eq:Wed4Aug093804CEST2021}) that covers all the cases
    simultaneously (but through a very bulky expression).}
  \label{fig:Fri6Aug173326CEST2021}
\end{figure*}

Now turning to Eq.~(\ref{eq:Mon11May2020131426BST}) with the
Hamiltonian~(\ref{eq:Mon11May2020131538BST}) complemented with a
detector~$\xi$, we obtain the time-resolved frequency-resolved
correlations for the coherently-driven 2LS. The general expression can
be obtained analytically but it becomes quite cumbersome and will
require the definition of a few auxiliary notations. It takes the form
of a sum of seven exponentials of~$\tau$ correcting the no-correlation
value (unity):
\begin{equation}
  \label{eq:Wed4Aug093804CEST2021}
  g^{(2)}_\sigma(\tau)=1+\sum_{i=1}^7\mathcal{G}_i^{(2)}\exp(-\gamma_i\tau)
\end{equation}
with coherence times
\begin{subequations}
  \label{eq:Wed4Aug094828CEST2021}
  \begin{align}
    \gamma_1&\equiv(3\gamma_\sigma+\gamma_\mathrm{M})/4\,,\\
    \gamma_2&\equiv(3\gamma_\sigma-\gamma_\mathrm{M})/4\,,\\
    \gamma_3&\equiv\Gamma/2\,,\\
    \gamma_4&\equiv\gamma_{{11}}/2\,,\\
    \gamma_5&\equiv(\gamma_{{23}}+\gamma_\mathrm{M})/4\,,\\
    \gamma_6&\equiv(\gamma_{{23}}-\gamma_\mathrm{M})/4\,,\\
    \gamma_7&\equiv\Gamma\,,
  \end{align}
\end{subequations}
where we have introduced the notations and
\begin{equation}
  \label{eq:Wed4Aug095911CEST2021}
  \gamma_{{ij}}\equiv i \Gamma + j\gamma_\sigma  
\end{equation}
i.e., $\gamma_{{11}}=\Gamma+\gamma_\sigma$ and
$\gamma_{{23}}=2\Gamma+3\gamma_\sigma$ (a bar means taking the
negative number).  The corresponding coefficients~$\mathcal{G}_i$ are
bulky and given in Appendix~\ref{sec:Wed4Aug100528CEST2021} as
Eqs.~(\ref{eq:Wed4Aug101509CEST2021}--\ref{eq:Wed4Aug144101CEST2021}),
but various limits of interest can be obtained in closed-form of size
reasonable enough to be featured as complete expressions, starting
with the zero-delay correlation for arbitrary driving intensities:
\begin{widetext}
  \begin{multline}
    \label{eq:Wed3Jun2020232311BST}
    g_\sigma^{(2)}(0)=
    \frac{\gamma_{{11}}(\gamma_{{01}}^2+4\Omega^2)
      (\gamma_{{11}}\gamma_{{12}}+8\Omega^2)
      [\gamma_{{11}}\gamma_{{21}}^2\gamma_{{31}}^2
      \gamma_{{12}}\gamma_{{32}}+4\gamma_{{10}}
      \gamma_{{31}}(17\gamma_{{10}}^3+
      29\gamma_{{10}}^2   
      \gamma_{{01}}+18\gamma_{{10}}\gamma_{{01}}^2+
      4\gamma_{{01}}^3)\Omega^2+48\gamma_{{10}}^2
      \gamma_{{21}}\Omega_0^4]}{\gamma_{{21}}
      \gamma_{{31}}(\gamma_{{11}}\gamma_{{21}}+
      4\Omega^2)(\gamma_{{31}}\gamma_{{32}}+8\Omega^2)
      (\gamma_{{11}}^2\gamma_{{12}}+4\gamma_{{10}}
      \Omega^2)^2}\,,
  \end{multline}
%
  with $\gamma_{{ij}}$ defined in
  Eq.~(\ref{eq:Wed4Aug095911CEST2021}), e.g.,
  $\gamma_{{31}}=3\Gamma+\gamma_\sigma$.  This was also obtained from
  the cascaded formalism in Ref.~\cite{lopezcarreno16a}, Eq.~(19b),
  and used to account for finite-driving departures in
  Ref.~\cite{hanschke20a} where this expressions provided an
  essentially exact fit to the raw data. The time-correlations in the
  low-driving, Heitler limit can be substantially simplified:
%
\begin{equation}
  \label{eq:Wed4Aug191616CEST2021}
  g^{(2)}_{\sigma,\mathrm{Heitler}}(\tau)=
  {e^{-(\Gamma+\gamma_\sigma)\tau}\left(\Gamma^2e^{\Gamma\tau/2}-\Gamma\gamma_\sigma e^{\gamma_\sigma\tau/2}-(\Gamma^2-\gamma_\sigma^2)e^{(\Gamma+\gamma_\sigma)\tau/2}\right)^2\over(\Gamma^2-\gamma_\sigma^2)^2}\,.
\end{equation}
with a simple overall behaviour of a monotonous loss of antibunching
due to frequency filtering given by:
\begin{equation}
  \label{eq:Mon11May2020190724BST}
  g^{(2)}_{\sigma,\mathrm{Heitler}}(0) =  \left (
    \frac{\gamma_\sigma}{\gamma_\sigma + \Gamma}
  \right )^2\,.
\end{equation}

%
The high-driving, Mollow regime, on the other hand, is more
complex. In particular, taking the limit~$\Omega_\sigma\to\infty$
sends the satellite peaks away and filtering will be limited to the
central peak in this case, which can only produce bunching, namely
\begin{equation}
  \label{eq:Wed4Aug192756CEST2021}
  g^{(2)}_{\sigma,\mathrm{Mollow}\atop 0\ll\Gamma\ll\Omega_\sigma}(\tau)=  
  1+2\gamma_\sigma{2\Gamma e^{-(\Gamma+\gamma_\sigma)\tau/2}-(\Gamma+\gamma_\sigma)e^{-\Gamma\tau}\over(\Gamma-\gamma_\sigma)(3\Gamma+\gamma_\sigma)}\,,
\end{equation}
with zero-delay coincidences
\begin{equation}
  \label{eq:Wed4Aug195200CEST2021}
  g^{(2)}_{\sigma,\mathrm{Mollow}\atop 0\ll\Gamma\ll\Omega_\sigma}(0)= {3(\Gamma+\gamma_\sigma)\over3\Gamma+\gamma_\sigma}\,.
\end{equation}
This corresponds to the central part of
Fig.~\ref{fig:Fri6Aug173326CEST2021}. The two other parts have to be
treated independently and retain the~$\Omega_\sigma$ dependence, to
yield:
\begin{subequations}
  \label{eq:Fri6Aug175436CEST2021}
  \begin{align}
    g^{(2)}_{\sigma,\mathrm{Mollow}\atop \Gamma\to0}(\tau)=&
 1+\left({2\gamma_\sigma^2\over\Omega_\sigma^2}+{4\Gamma\gamma_{2\overline{1}}\over\gamma_\sigma^2}\right)e^{-(\Gamma+\gamma_\sigma)\tau/2}+{3\gamma_\sigma^8-8\Gamma\gamma_\sigma^4\gamma_{11}\Omega_\sigma^2+16\Gamma^2(\Gamma\gamma_{5\overline{1}}+\gamma_\sigma^2)\Omega_\sigma^4\over8\Gamma^2\gamma_\sigma^2\Omega_\sigma^4}e^{-\Gamma\tau}
\label{eq:Fri6Aug182059CEST2021}\\
    g^{(2)}_{\sigma,\mathrm{Mollow}\atop \Gamma\gg\Omega_\sigma}(\tau)=&1
+{8(\Gamma/\Omega_\sigma)^2(10+(\Gamma/\Omega_\sigma)^2)\over(4+(\Gamma/\Omega_\sigma)^2)(8+(\Gamma/\Omega_\sigma)^2)^2}e^{-\Gamma\tau}
                                                                         -{4(\Gamma/\Omega_\sigma)^2(16+(\Gamma/\Omega_\sigma)^2)\over(4+(\Gamma/\Omega_\sigma)^2)(8+(\Gamma/\Omega_\sigma)^2)^2}e^{-(\gamma_\sigma+\Gamma)\tau/2}\nonumber\\
                                                           &+{4(\Gamma/\Omega_\sigma)^2\big((16+(\Gamma/\Omega_\sigma)^2\cos(2\Omega_\sigma\tau)+(\Gamma/\Omega_\sigma)(10+(\Gamma/\Omega_\sigma)^2)\sin(2\Omega_\sigma\tau)\big)\over(4+(\Gamma/\Omega_\sigma)^2)(8+(\Gamma/\Omega_\sigma)^2)^2}e^{-(3\gamma_\sigma+2\Gamma)\tau/4}\label{eq:Fri6Aug182129CEST2021}\\
        &-{(\Gamma/\Omega_\sigma)^2(16+(\Gamma/\Omega_\sigma)^2)\cos(2\Omega_\sigma\tau)\over(8+(\Gamma/\Omega_\sigma)^2)^2}e^{-3\gamma_\sigma\tau/4}\,,\nonumber
  \end{align}
\end{subequations}
with corresponding zero-delay correlations:
\begin{subequations}
  \label{eq:Fri6Aug173936CEST2021}
  \begin{align}
      g^{(2)}_{\sigma,\mathrm{Mollow}\atop \Gamma\to0}(0)&={{\gamma_{41}\gamma_\sigma^9+12\Gamma\gamma_{21}\gamma_\sigma^6\Omega_\sigma^2}{-16\Gamma^2\gamma_\sigma^3\gamma_{14,\overline{5}}\Omega_\sigma^4-192\Gamma^3\gamma_{2\overline{1}}\Omega_\sigma^6}\over\gamma_\sigma(\gamma_\sigma^3+4\Gamma\Omega_\sigma^2)^3}\,,\label{eq:Fri6Aug181717CEST2021}\\
      g^{(2)}_{\sigma,\mathrm{Mollow}\atop \Gamma\gg\Omega_\sigma}(0)&={8(2+(\Gamma/\Omega_\sigma)^2)(16+(\Gamma/\Omega_\sigma)^2)\over(4+(\Gamma/\Omega_\sigma)^2)(8+(\Gamma/\Omega_\sigma)^2)^2}\,,\label{eq:Fri6Aug181731CEST2021}
  \end{align}
\end{subequations}
which correspond to the left and right parts of
Fig.~\ref{fig:Fri6Aug173326CEST2021}, respectively. The simplest
formula we can find that unites all these behaviours is the one that
is valid for all~$\Omega_\sigma$ anyway, namely,
Eq.~(\ref{eq:Wed3Jun2020232311BST}) to describe the three limits of
Eqs.~(\ref{eq:Fri6Aug181717CEST2021}),
(\ref{eq:Wed4Aug195200CEST2021}) and~(\ref{eq:Fri6Aug181731CEST2021})
that together reconstruct the curve in
Fig.~\ref{fig:Fri6Aug173326CEST2021}(a), and, for the~$\tau$
dependence, Eq.~(\ref{eq:Wed4Aug093804CEST2021}) along
with~(\ref{eq:Wed4Aug094828CEST2021})
and~(\ref{eq:Wed4Aug101509CEST2021}--\ref{eq:Wed4Aug144101CEST2021})
for the corresponding limits of Eqs.~(\ref{eq:Fri6Aug182059CEST2021}),
(\ref{eq:Wed4Aug192756CEST2021})
and~(\ref{eq:Fri6Aug182129CEST2021}). Some of the above results are
quite heavy but that only illustrates how rich and complex is the
seemingly simple and basic problem of filtering a two-level system and
the sort of complexity needed to embody all these various regimes and
behaviours in a single analytical expression. 
\end{widetext}
One can also check that these results recover those of the unfiltered
case in the limit~$\Gamma\to\infty$, namely, both
Eqs.~(\ref{eq:Wed4Aug093804CEST2021})
and~(\ref{eq:Fri6Aug182129CEST2021}) recover
Eq.~(\ref{eq:Thu23Apr2020230012BST}),
while~Eq.~(\ref{eq:Fri6Aug181731CEST2021}) simply recovers the perfect
antibunching
$g_{\sigma,\Omega\gg\gamma_\sigma,\Gamma\to\infty}^{(2)}(0) = 0$,
provided, again, that the limit be taken in the correct order (the
filtering should be larger than the Mollow splitting). The situation
is, again, simpler in the Heitler regime where also
Eqs~(\ref{eq:Wed4Aug191616CEST2021}) recovers
Eq.~(\ref{eq:Thu23Apr2020230640BST}) and
Eq.~(\ref{eq:Mon11May2020190724BST}) going to~0, much faster than the
Mollow antibunching, as shown in Fig.~\ref{fig:Fri6Aug173326CEST2021}.

We show some temporal correlations for the coherently-driven SPS in
Figure~\ref{fig:Mon11May2020173914BST}(a), as a function of time and
the linewidth of the detector, for a 2LS with a
driving~$\Omega_\sigma=2\gamma_\sigma$, thus well within the Mollow
regime of excitation (the loss of antibunching in the Heitler regime
has been recently studied in detail both theoretically, along the
lines of the current text, and
experimentally~\cite{phillips20a,hanschke20a}). On the upper part,
where the detector is colorblind, the correlations are given by
Eq.~(\ref{eq:Thu23Apr2020230012BST}) and the Rabi oscillations induced
by the laser have the maximum visibility~\cite{makhonin14a}.
\begin{figure}[t]
  \includegraphics[width=\linewidth]{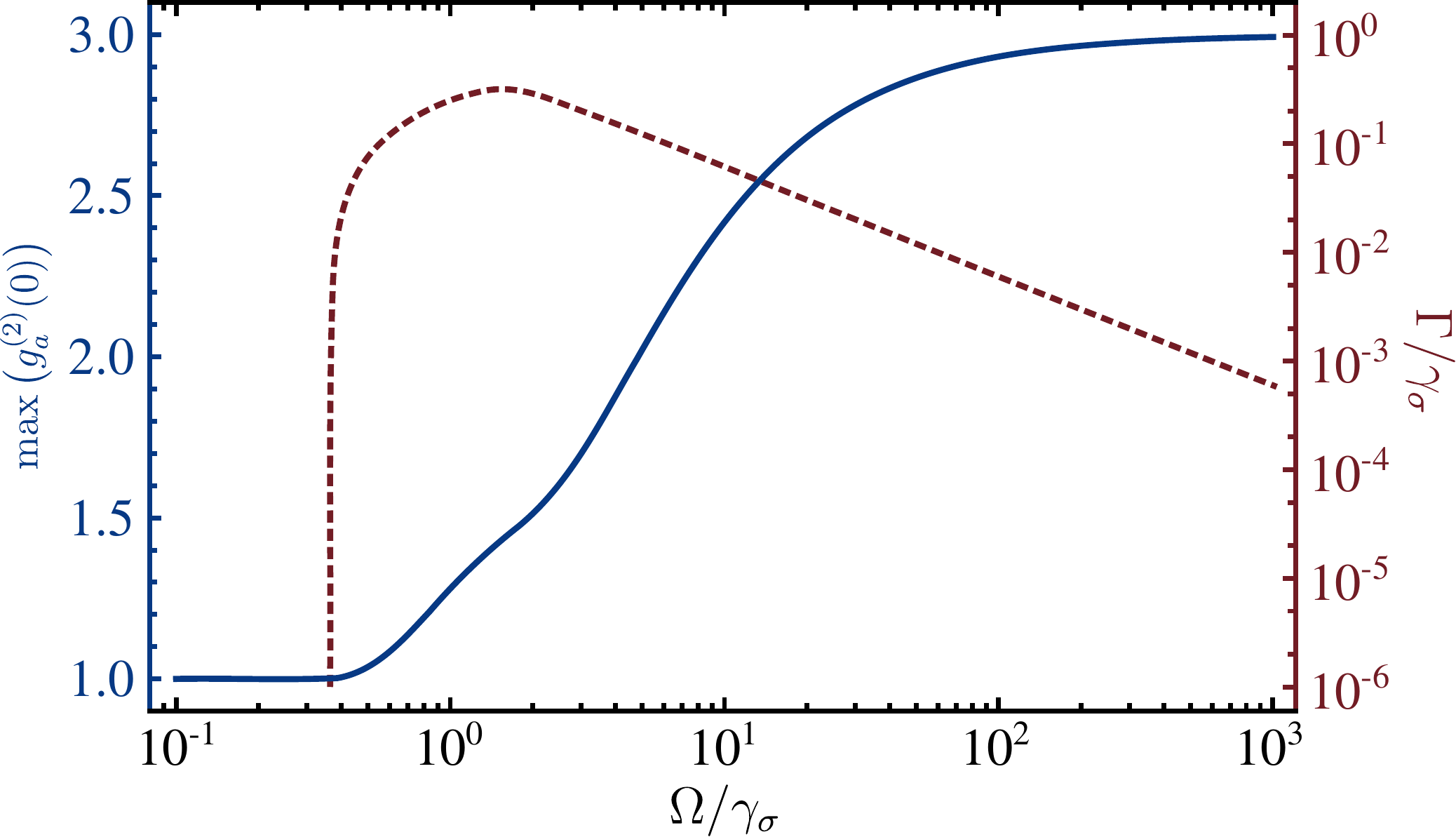}   
  \caption{(solid, blue) Maximum value that the zero-delay
    correlation~$g^{(2)}_\sigma(0)$ of the coherently-driven 2LS can
    take for a given driving intensity~$\Omega$.  At every intensity
    of the driving, the filter's linewidth $\Gamma$ has been taken to
    maximize~$g^{(2)}_\sigma(0)$ and is shown in dashed, red with
    values on the right-axis. This shows that a coherently-driven 2LS
    can always fail to display antibunching depending on the
    filtering.  At $\Omega/\gamma_\sigma\lessapprox0.4$, the filter's
    width gets too small to still be featured on the plot but
    theoretically exists and still remain larger than the~$\delta$
    function (that has no width) as to produce uncorrelated
    photons. It would produce bunching otherwise. The qualitative
    change at about~$\Omega/\gamma_\sigma \sim 2$ occurs when the
    three peaks of the Mollow triplet start to be distinguished.}
  \label{fig:Fri5Jun2020225354BST} 
\end{figure}
As the linewidth of the detector becomes commensurable with the
linewidth of the 2LS, and therefore starts filtering mostly the
central peak of the Mollow triplet, the oscillations are softened and
vanish completely when the detector filters within a region narrower
than the linewidth of the 2LS. Panel~(b) shows a series of cuts for
several values of~$\Gamma/\gamma_\sigma$, marked in~(a) by dashed
lines. The transition from perfect detection in time to perfect
detection in frequency can pass through an intermediate regime in
which the correlations are bunched. This is clear when focusing on the
zero-delay correlations~(\ref{eq:Wed3Jun2020232311BST}), as shown in
Fig.~\ref{fig:Mon11May2020173914BST}(c) as a function of the linewidth
of the detector for several intensities of the driving. The
low-driving case has been observed experimentally by two independent
groups~\cite{phillips20a,hanschke20a}. One can observe striking
differences with respect to the case of incoherent
driving~(\ref{fig:Fri8May2020215520BST}(c)): first, the limit of
vanishing filtering width leads to randomization rather than to
thermalization, thereby behaving like a time-jitter rather than
thermal filtering. Such a behaviour is a manifestation of the type of
driving and the approximation made to describe the laser as
a~$\delta$-function, which has exactly zero linewidth. Because of this
approximation, one can never filter the photons from within the
emission line of the laser, and therefore the thermalization does not
appear. However, turning to more sophisticated models of the laser,
e.g., the one atom laser~\cite{mu92a}, then on reaches a
thermalization in the limit~$\Gamma \rightarrow 0$.  The widely used
approximation of the laser as a~$\delta$-function has nevertheless
been shown to be good enough to account for experimental observations
several times and under different conditions: mapping the two- and
three-photon correlation landscape of the 2LS~\cite{peiris15a,
  nieves18a} (in agreement with the theoretical
predictions~\cite{gonzaleztudela13a, lopezcarreno17a}) and recently by
measuring the effect of the filter~\cite{hanschke20a} in perturbing
the balance between the quantum emission and the laser itself in a
self-homodyning picture~\cite{zubizarretacasalengua20a}.  In this
case, when the linewidth of the filtering is not vanishing, but still
remains below the natural linewidth of the 2LS, the correlations
become bunched as the intensity of the driving increases. This brings
us to the second main departure from the incoherent case, namely, the
non-monotonous evolution of~$g_\sigma^{(2)}(0)$ even if it would
thermalize at vanishing~$\Gamma$. One can find a region of
superbunching with more bunched photons emitted by the 2LS than
thermal light itself. The maximum bunching is $g_\sigma^{(2)}(0)=3$,
as shown in Fig.~\ref{fig:Fri5Jun2020225354BST} and is obtained in the
limit of high driving and vanishing filtering. This is realized after
the observed qualitative change
at~$\Omega_\sigma/\gamma_\sigma \geq 1/\sqrt{8}$, when the intensity
of the driving is such that the emission spectrum of the 2LS is given
by the sum of three Lorentzians~\cite{lopezcarreno16a}. Satisfying the
latter condition implies entering into the Mollow triplet regime, and
filtering below the natural linewidth of the 2LS means that one is
filtering the photons coming from the central peak. The correlations
from the photons coming from this region have been predicted to
display bunching~\cite{schrama92a, nienhuis93a, gonzaleztudela13a},
which has since been observed experimentally~\cite{peiris15a}. The
bunching also observed before that point can be explained by the
self-homodyning picture~\cite{zubizarretacasalengua20b}, which we do
not discuss further as this relate to bunching rather than
antibunching that comes with specificities of its own, but this serves
to illustrate how the regime of driving result in very different
dynamics for the antibunching of the emitter. In particular, the loss
of antinbunching is so serious for the coherently-driven 2LS that it
complete for all driving intensities, if the filtering is too
stringent.

\section*{Conclusions and perspectives}

We have described a variety of mechanisms---in our understanding the
most common and important ones---that lead to a reduction of
antibunching with a particular focus on that produced by a two-level
system under various regimes of excitations. Much of our results
either apply directly to other cases, including both other systems and
other types of photon statistics, or could be adapted to systems of
interest following the above techniques. For instance, it would be
instructive to adapt this analysis to the current record-holder of
antibunched emission~\cite{schweickert18a, hanschke18a} and track
whether the origin of the remaining imperfection lies in the factors
analyzed here or in the source itself (i.e., residual
re-excitation). We surmise that a perfect antibunched source should
not have a Lorentzian profile but feature a standard deviation so as
to avoid spurious coincidences from the finite-bandwidth of the
detector. This, in itself, however constitutes a different
problem. 

Even the 2LS system that we have studied above could be analyzed
further, e.g., to higher photon-orders, out of resonance with the
filter, including other processes such as pure dephasing or
phonon-induced dephasing, etc.  One could also extend the discussion
to other mechanisms that spoil antibunching (some, such as the gravity
peak of a streak camera whereby one photon spreads over several pixels
of the detectors, are covered in Ref.~\cite{blancathesis}) or combine
those that we have discussed above, from either loss by external noise
(Section~\ref{sec:Fri30Jul190600CEST2021}), by time uncertainty (or
time jitter) (Sec.~\ref{sec:Fri30Jul190612CEST2021}) and/or by
frequency filtering (Sec.~\ref{sec:Fri30Jul190644CEST2021}). There are
indeed more mechanisms than one would actually want to account for,
which lead to a loss of antibunching, and most of them do not
constitute in a simple exponential damping of the perfect antibunching
of the source. While one particular mechanism (frequency filtering)
has been recently investigated in-depth for one of the regimes
considered above (Heitler regime)~\cite{phillips20a,hanschke20a},
there remains much to confirm experimentally. In particular, the
transition frmo antibunching to bunching for the incoherently-driven
two-level system, tending to thermalization ruled by the filter alone
and passing by an imperfectly uncorrelated state which exhibits a
small supression of photon pairs within the emitter's coherence time
when exact coincidences are those of a coherent state
($g^{(2)}(0)=1$), remain to be experimentally demonstrated, a
considerable omission for what is arguably the most widely studied
quantum optical emitter. The level of agreement with the theory would
allow to assess how valid is the two-level picture for the emitter in
question, how well-understood is the mechanism leading to its loss of
antibunching as well as, at a more fundamental level, how accurate is
the theory of frequency-resolved photon correlations, which
constitutes in itself a basic aspect of the theory of photodetection.

\begin{acknowledgements}
  We thank D. Sanvitto and C. Sanchez Mu\~noz for comments and
  fruitful discussions over the years in this long-delayed manuscript,
  as well as J. Warner and D. Weston for their assistance in compiling
  the literature of small-$g^{(2)}(0)$ values.
\end{acknowledgements}

\onecolumngrid
\appendix

\section*{Appendix: Analytical expresssions}

The appendix gives the closed-form analytical expressions which are
needed, discussed and/or plotted in the text.

\section{Time Jitter}

\subsection{Heaviside function}
\label{eq:Tue17Nov093047CET2020}

The case of the Heaviside function
Eq.~(\ref{eq:Fri24Apr2020111614BST}) yields for
Eq.~(\ref{eq:Fri24Apr2020113650BST})
\begin{multline}
  \label{sec:Fri24Apr2020223850BST}
  g^{(2)}_\Gamma (\tau) = \Gamma \int_{\tau}^{\tau+\frac{1}{\Gamma}}
  \!\!dx\,g^{(2)}(x) 
  [1-\Gamma (x-\tau)] + \theta \left(\tau- \frac{1}{\Gamma} \right) \Gamma
  \int_{\tau-\frac{1}{\Gamma}}^\tau \!\!dx\,g^{(2)}(x) [1+\Gamma (x-\tau)] 
  \\
  {}+\theta \left( \frac{1}{\Gamma} -
    \tau \right) \Gamma \left \lbrace  \int_0^{\frac{1}{\Gamma}-\tau} \!\!dx\,
  g^{(2)}(x) \left [1-\Gamma(x+\tau) \right] +{} \right.
  + \left. \int_0^{\tau} \!\!dx\, 
  g^{(2)}(x) \left [1+\Gamma(x-\tau) \right] \right \rbrace\,,
\end{multline}
%
providing the photon
correlations from an incoherently driven 2LS with time jitter:
%
\begin{align}
  \label{eq:Tue17Nov093639CET2020}
  g^{(2)}_{\Gamma,P_\sigma} (\tau) = 1 &-{\Gamma^2\over\Gamma_\sigma^2}e^{-\Gamma_\sigma(\tau+1/\Gamma)}(1-e^{\Gamma_\sigma/\Gamma})^2\theta\left(\tau-{1\over\Gamma}\right)\\
  {}&-{2\Gamma^2\over\Gamma_\sigma^2}\left\{e^{-\Gamma_\sigma/\Gamma}\cosh(\Gamma_\sigma\tau)-e^{-\Gamma_\sigma\tau}+\Gamma_\sigma({1\over\Gamma}-\tau)\right\}\theta\left({1\over\Gamma}-\tau\right)
\,,
\end{align}
with the $\tau\ge1/\Gamma$ behaviour on the first line and
$\tau\le1/\Gamma$ on the second line, while the corresponding
expression for the coherently driven 2LS is
\begin{multline}
  \label{eq:Tue28Apr2020213841BST}
  g^{(2)}_{\Gamma,\Omega}(\tau) = 1 -
  \frac{32\Gamma^2 e^{-3\gamma_\sigma \tau/4}}{R_\sigma(R_\sigma^2
    +9\gamma_\sigma^2)^2} \left [ 9\gamma_\sigma
    (R_\sigma^2-3\gamma_\sigma^2)  \sin\left(
      \frac{R_\sigma \tau}{4}\right) + R_\sigma(R_\sigma^2 - 27
    \gamma_\sigma^2) \cos \left(
    \frac{R_\sigma \tau}{4}\right) \right] +{}\\
  {}+  \frac{16\Gamma e^{-3\gamma_\sigma \tau/4}
      e^{-3\gamma_\sigma/(4\Gamma)}}{R_\sigma(R_\sigma^2 
    +9\gamma_\sigma^2)^2}\left\lbrace
    9\gamma_\sigma(R_\sigma^2-3\gamma_\sigma^2) \sin \left[
      \frac{R_\sigma(1+\Gamma \tau)}{4\Gamma}\right] +
    R_\sigma(R_\sigma^2-27\gamma_\sigma^2) \cos\left[
      \frac{R_\sigma(1+\Gamma \tau)}{4\Gamma}\right] \right \rbrace +{}\\ 
  {}+\frac{16\Gamma^2(R_\sigma^2-27\gamma_\sigma^2)}{(R_\sigma^2+ 9
    \gamma_\sigma^2)^2} \cos \left[ \frac{R_\sigma (1-\Gamma \tau)}{4\Gamma}
  \right] e^{-\frac{3\gamma_\sigma|1-\Gamma \tau|}{4\Gamma}}
 -\frac{48 \Gamma \gamma_\sigma
    (1-\Gamma \tau)}{R_\sigma^2+9\gamma_\sigma^2} \theta \left(
    \frac{1}{\Gamma} - \tau \right) +{}\\
  {}+\frac{144\Gamma^2\gamma_\sigma
    (R_\sigma^2-3\gamma_\sigma^2)}{R_\sigma(R_\sigma^2+ 9 
    \gamma_\sigma^2)^2} \sin \left[ \frac{R_\sigma (1-\Gamma \tau)}{4\Gamma}
  \right] \left [ e^{-\frac{3\gamma_\sigma(1-\Gamma \tau)}{4\Gamma}}
    \theta \left( 
    \frac{1}{\Gamma} - \tau \right) -
  e^{\frac{3\gamma_\sigma(1-\Gamma \tau)}{4\Gamma}} 
    \theta \left( \tau - \frac{1}{\Gamma} \right)   \right]\,.
\end{multline}

\subsection{Exponential function}
\label{sec:Thu23Apr2020212320BST}

In this case, the correlations with time jitter is given by
\begin{equation}
  \label{eq:Tue28Apr2020131623BST}
  g^{(2)}_\Gamma (\tau) = \Gamma \cosh(\Gamma \tau) \int_0^\infty g^{(2)}(x) e^{-\Gamma x}\,dx - \Gamma \int_0^\tau\,g^{(2)}(x) \sinh\left[ \Gamma(\tau-x)\right]\,dx\,.
\end{equation}
In the case of the incoherently driven 2LS, the correlations with time
jitter become
\begin{equation}
  \label{eq:Tue28Apr2020132411BST}
  g^{(2)}_{\Gamma,P_\sigma} (\tau) = 1- {1\over 1-(\Gamma_\sigma/\Gamma)^2}
  \left(  e^{-\Gamma_\sigma\tau} - (\Gamma_\sigma/\Gamma) e^{-\Gamma
      \tau}  \right)\,,  
\end{equation}
with limit
$g^{(2)}_{\Gamma,P_\sigma}
(\tau)=1-e^{-\Gamma_\sigma\tau}(1+\Gamma_\sigma\tau)/2$ when
$\Gamma=\Gamma_\sigma$, whereas the correlations of the coherently
driven 2LS become
\begin{equation}
  \label{eq:Tue28Apr2020135027BST}
  g^{(2)}_{\Gamma,\Omega} (\tau) = 1-
  \frac{j_{1,\Omega}}{N_+N_-}e^{-\Gamma \tau} -
  \frac{j_{2,\Omega}}{R_\sigma N_+N_-}e^{-3\gamma_\sigma \tau/2}\,,
\end{equation}
where~$N_\pm = R_\sigma^2 + (4\Gamma \pm 3\gamma_\sigma)^2$ and we
have defined the functions
\begin{subequations}
  \begin{equation}
    \label{eq:Tue28Apr2020135905BST}
    j_{1,\Omega}= 24 \Gamma \gamma_\sigma (R_\sigma^2
    +9\gamma_\sigma^2)\,,
  \end{equation}
  \begin{equation}
    \label{eq:Tue28Apr2020140318BST}
    j_{2,\Omega} = 16 R_\sigma \Gamma^2 (R_\sigma^2 +16\Gamma^2 -
    27\gamma_\sigma^2) \cos(R_\sigma\tau/4) + 48\Gamma^2\gamma_\sigma (3R_\sigma^2
    +16\Gamma^2-9\gamma_\sigma^2) \sin(R_\sigma\tau /4)\,.
  \end{equation}
\end{subequations}

\subsection{Double-exponential}

 With this function, the expression for the time jitter correlations
 are given by
\begin{multline}
  \label{eq:Fri24Apr2020232245BST}
  g^{(2)}_\Gamma(\tau) = \Gamma \int_0^\infty dx\,g^{(2)}(x)e^{-2\Gamma
  x}  \left [ (1+2x\Gamma) \cosh(2\Gamma\tau)\right. -{}\\ \left. 2\Gamma\tau
  \sinh(2\Gamma\tau ) \right ] - \Gamma \int_0^\tau dx\,g^{(2)}(x) \left
  \lbrace \sinh[2\Gamma (\tau-x)]-{}\right.\\
  \left. {}-2\Gamma(\tau -x) \cosh[2\Gamma(\tau-x)] \right\rbrace\,.
\end{multline}
In this case, the correlations of the incoherently driven 2LS, with
bare correlations given in Eq.~(\ref{eq:Thu23Apr2020225637BST}),
become
\begin{equation}
  \label{eq:Tue17Nov110858CET2020}
  g^{(2)}_{\Gamma,P_\sigma}(\tau) = 1 -
  \frac{16}{\Delta^2}e^{-\Gamma_\sigma \tau} 
  +\frac{(\Gamma_\sigma/\Gamma)\big(8+\Delta(1+2\Gamma\tau)\big)}{\Delta^2}
  e^{-2\Gamma\tau}\,, 
\end{equation}
where we have used the
notation~$\Delta\equiv 4-(\Gamma_\sigma/\Gamma)^2$. The counterpart
for the coherent excitation is given by
\begin{equation}
  \label{eq:Sat25Apr2020192637BST}
  g_{\Gamma,\Omega}^{(2)}(\tau) = 1 -
    \frac{g_{1,\Omega}}{\mathcal{N}_\Omega} e^{-3\gamma_\sigma \tau/4 } -
    \frac{g_{2,\Omega}}{\mathcal{N}_\Omega} e^{-2\Gamma \tau}\,,
\end{equation}
where we have introduced the functions~$g_{1,\Omega}$, $g_{2,\Omega}$
and~$\mathcal{N}_\Omega$ as follows
\begin{subequations}
  \begin{multline}
    g_{1,\Omega} = 4096 \Gamma^4 \left \lbrace R_\sigma
      \left[(R_\sigma^2+64 \Gamma^2)^2 -18(5R_\sigma^2
        +192 \Gamma^2)\gamma_\sigma^2 + 405
        \gamma_\sigma^4 \right] \cos(R_\sigma\tau/4) +{}
    \right.\\ 
    \left. {}+3\gamma_\sigma \left[5R_\sigma^4 + 6R_\sigma^2
        (64\Gamma^2 -15\gamma_\sigma^2) + (64\Gamma^2 -
        9\gamma_\sigma^2)^2 \right] \sin(R_\sigma
      \tau/4) \right \rbrace\,, 
  \end{multline}
  \begin{multline}
    \label{eq:Sat25Apr2020193458BST}
    g_{2,\Omega} = 24 R_\sigma \Gamma \gamma_\sigma (R_\sigma^2 + 9
    \gamma_\sigma^2) + \left \lbrace R_\sigma^4 +384 R_\sigma^2
      \Gamma^2 +20480 \Gamma^4- 3456 \Gamma^2 \gamma_\sigma^2 +81
      \gamma_\sigma^4 +{} \right.\\
    \left.{}+ 2\Gamma [R_\sigma^2 + (8\Gamma-3\gamma_\sigma)^2]
      [R_\sigma^2 + (8\Gamma+3\gamma_\sigma)^2]\tau \right \rbrace\,,
  \end{multline}
  \begin{equation}
    \label{eq:Sat25Apr2020193513BST}
    \mathcal{N}_\Omega = R_\sigma \left [R_\sigma^4 + 2R_\sigma^2
      (64\Gamma^2 + 9\gamma_\sigma^2) + (64\Gamma^2 -
      9\gamma_\sigma^2)^2   \right]\,.
  \end{equation}
\end{subequations}

\subsection{Gaussian}

\begin{equation}
  \label{eq:Sat25Apr2020200043BST}
  g_{\Gamma}^{(2)}(\tau) = \frac{\Gamma}{2\sqrt{\pi}} \int_0^\infty
  g^{(2)}(x) \left \lbrace e^{-[(\tau+x)\Gamma/2]^2}
    +e^{-[(\tau-x)\Gamma/2]^2}\right \rbrace\,dx\,.  
\end{equation}

The correlation with time uncertainty for the incoherently driven 2LS
becomes
\begin{equation}
  \label{eq:Sat25Apr2020201941BST}
  g_{\Gamma,P_\sigma}^{(2)}(\tau) = 1 -
  \frac{e^{(\Gamma_\sigma/\Gamma)^2}}{2} \left \lbrace
    e^{-\Gamma_\sigma \tau} \mathrm{Erfc}(\tau_-) + e^{\Gamma_\sigma
      \tau} \mathrm{Erfc}(\tau_+)  \right \rbrace\,,
\end{equation}
where~$\mathrm{Erfc(\tau)}$ is the complementary error function and we
have defined
\begin{equation*}
  \tau_\pm = \frac{\Gamma_\sigma}{\Gamma} \pm \frac{\Gamma \tau}{2}\,.
\end{equation*}
The counterpart for coherent excitation has a more complicated
structure
\begin{equation}
  \label{eq:Mon27Apr2020205804BST}
  g_{\Gamma,\Omega}^{(2)}(\tau) = 1- \frac{\Delta_-}{i R_\sigma}
  e^{(\Delta_+/\Gamma)^2} h_{1,\Omega} - \frac{\Delta_+}{iR_\sigma}
  e^{(\Delta_-/\Gamma)^2} h_{2,\Omega}\,,
\end{equation}
where we have introduced the functions~$h_{1,\Omega}$, $h_{2,\Omega}$
and the parameters~$\Delta_\pm$, defined as
\begin{subequations}
  \begin{equation}
    \label{eq:Mon27Apr2020211706BST}
    h_{1,\Omega} = 2 \cosh(\Delta_+ \tau)
    +\mathrm{Erf}(\lambda_{1,+}) e^{\Delta_+ \tau} +
    \mathrm{Erf}(\lambda_{1,-}) e^{-\Delta_+ \tau}\,,
  \end{equation}
  \begin{equation}
    \label{eq:Mon27Apr2020211713BST}
    h_{2,\Omega} =2 \cosh(\Delta_- \tau)
    -\mathrm{Erf}(\lambda_{2,+}) e^{\Delta_- \tau} -
    \mathrm{Erf}(\lambda_{2,-}) e^{-\Delta_- \tau}\,,
  \end{equation}
\end{subequations}
where we have used~$\Delta_\pm = (iR_\sigma \pm 3 \gamma_\sigma)/4$,
$\lambda_{1,\pm}= \Delta_+/\Gamma \pm \Gamma \tau/2$,
$\lambda_{2,\pm}= \Delta_-/\Gamma \pm \Gamma \tau/2$ and~$R_\sigma$ is
as defined in Eq.~(\ref{eq:Thu23Apr2020230012BST}).

\section{Frequency Filtering}
\label{sec:Wed4Aug100528CEST2021}

The seven coefficients~$\mathcal{G}_i$ which, together with the
coherence times~(\ref{eq:Wed4Aug094828CEST2021}), yield the general
two-photon auto-correlation function~$g^{(2)}_\sigma(\tau)$ for the
coherently driven 2LS according to
Eq.~(\ref{eq:Wed4Aug093804CEST2021}) are given below. They also
consist of intricate combinations of the various rates involved, this
time also involving subtractions, so that we upgrade
Eq.~(\ref{eq:Wed4Aug095911CEST2021}) 

\begin{equation}
  \label{eq:Wed4Aug101509CEST2021}
    \mathcal{G}_1\equiv
    {
    \splitfrac{512\Gamma^2\gamma_{11}\Omega_\sigma^2(\gamma_{11}\gamma_{12}+16\Omega_\sigma^2)(\Gamma\gamma_{1\overline{2}}\gamma_{1\overline{1}}(\gamma_\mathrm{M}+\gamma_\sigma)+8[14\Gamma^2+2\gamma_\sigma(\gamma_\mathrm{M}+\gamma_\sigma)-\Gamma(7\gamma_\mathrm{M}+17\gamma_\sigma)]\Omega_\sigma^2-512\Omega_\sigma^4)}
    {\kern.2cm\times(\gamma_{11}\gamma_{12}\gamma_{21}(\Gamma(\gamma_\mathrm{M}-3\gamma_\sigma)+2\gamma_\sigma(\gamma_\mathrm{M}-\gamma_\sigma))+8[8\Gamma^3+32\Gamma\gamma_\sigma^2-2\gamma_\sigma^2(\gamma_\mathrm{M}-7\gamma_\sigma)+\Gamma^2(\gamma_\mathrm{M}+25\gamma_\sigma)]\Omega_\sigma^2+256\Gamma\Omega_\sigma^4)}
    \over
    \gamma_\mathrm{M}(\gamma_\mathrm{M}-\Gamma)(\gamma_\mathrm{M}-\gamma_\sigma)(\gamma_\mathrm{M}+\gamma_\sigma)^2(\gamma_\mathrm{M}+\gamma_\sigma-2\Gamma)(\gamma_\mathrm{M}+3\gamma_\sigma-4\Gamma)(\gamma_\mathrm{M}+3\gamma_\sigma-2\Gamma)(\gamma_{11}\gamma_{21}+8\Omega_\sigma^2)(\gamma_{11}^2\gamma_{12}+8\Gamma\Omega_\sigma^2)^2
    }
\end{equation}

\begin{equation}
  \label{eq:Wed4Aug184040CEST2021}
  \mathcal{G}_2\equiv\mathcal{G}_1\ \hbox{with } \gamma_\mathrm{M}\leftrightarrow-\gamma_\mathrm{M}
\end{equation}

\begin{equation}
  \label{eq:Wed4Aug121211CEST2021}
  \mathcal{G}_3\equiv
  2{256\Gamma\gamma_{11}\gamma_\sigma\Omega_\sigma^2(\gamma_{11}\gamma_{12}+16\Omega_\sigma^2)(\gamma_{11}^2\gamma_{12}\gamma_{1\overline{2}}\gamma_{21}^2+8\gamma_{11}(9\Gamma^3+20\Gamma\gamma_\sigma\gamma_{11}+8\gamma_\sigma^3)\Omega_\sigma^2+128\Gamma^2\Omega_\sigma^4)
    \over
    \gamma_{11}^2\gamma_{21}(\gamma_\mathrm{M}^2-\gamma_\sigma^2)(\gamma_\mathrm{M}^2-\gamma_{2\overline{3}}^2)(\gamma_{11}\gamma_{21}+8\Omega_\sigma^2)(\gamma_{12}+8\Gamma\Omega_\sigma^2)^2}
\end{equation}

\begin{equation}
  \label{eq:Wed4Aug130157CEST2021}
  \mathcal{G}_4\equiv
2{2\Gamma^3(\gamma_\sigma^2+8\Omega_\sigma^2)(\gamma_{11}\gamma_{12}+16\Omega_\sigma^2)(\gamma_{11}^2\gamma_{12}^2\gamma_{31}+48\Gamma\gamma_{11}^2\Omega_\sigma^2-256\gamma_\sigma\Omega_\sigma^4)
\over
\gamma_{1\overline{1}}\gamma_{31}(\gamma_\mathrm{M}^2-\gamma_{2\overline{1}}^2)(\gamma_{11}\gamma_{21}+8\Omega_\sigma^2)(\gamma_{11}^2\gamma_{12}+8\Gamma\Omega_\sigma^2)^2
  }
\end{equation}

\begin{equation}
\label{eq:Wed4Aug190150CEST2021}
\mathcal{G}_5\equiv2\frac{
\begin{multlined}
-1024\Gamma^2\gamma_{11}\Omega_\sigma^4\Big[\gamma_{11}^2\gamma_{12}^2\gamma_{21}\gamma_{31}\gamma_{32}\big(2\Gamma^2(\gamma_\mathrm{M}-3\gamma_\sigma)-\Gamma\gamma_\sigma(\gamma_\mathrm{M}+3\gamma_\sigma)-2\gamma_\sigma^2(\gamma_\mathrm{M}-\gamma_\sigma)\big)+{}\\
{}+8\gamma_{11}\gamma_{12}\Omega_\sigma^2\Big\{-108\Gamma^6+\Gamma^5(215\gamma_\mathrm{M}-1203\gamma_\sigma)+3\Gamma^4\gamma_\sigma(239\gamma_\mathrm{M}-1081\gamma_\sigma)+\Gamma^3\gamma_\sigma^2(1051\gamma_\mathrm{M}-3947\gamma_\sigma)+{}\\
\hskip2.57cm{}+\Gamma^2\gamma_\sigma^3(803\gamma_\mathrm{M}-2465\gamma_\sigma)+10\Gamma\gamma_\sigma^4(31\gamma_\mathrm{M}-77\gamma_\sigma)+48\gamma_\sigma^5(\gamma_\mathrm{M}+2\gamma_\sigma)\Big\}+{}\\
{}+128\Omega_\sigma^4\Big\{
6\Gamma^6+\Gamma^5(131\gamma_\mathrm{M}-227\gamma_\sigma)+\Gamma^4\gamma_\sigma(546\gamma_\mathrm{M}-776\gamma_\sigma^2)+\Gamma^3\gamma_\sigma^2(889\gamma_\mathrm{M}-933\gamma_\sigma)+{}\\
+\Gamma^2\gamma_\sigma^3(724\gamma_\mathrm{M}-488\gamma_\sigma)+\Gamma\gamma_\sigma^4(296\gamma_\mathrm{M}-96\gamma_\sigma)+48\gamma_\mathrm{M}\gamma_\sigma^5
\Big\}+{}\\
{}+2048\Omega_\sigma^6\Big\{74\Gamma^4+2\Gamma^3(6\gamma_\mathrm{M}+109\gamma_\sigma)+5\Gamma^2\gamma_\sigma(3\gamma_\mathrm{M}+61\gamma_\sigma)+2\Gamma\gamma_\sigma^2(-\gamma_\mathrm{M}+103\gamma_\sigma)-4\gamma_\sigma^3(\gamma_\mathrm{M}-13\gamma_\sigma)\Big\}+{}\\
{}+131072\Gamma\gamma_{21}\Omega_\sigma^8\Big]
\end{multlined}
}{\gamma_{21}\gamma_\mathrm{M}(\gamma_\mathrm{M}+\Gamma)(\gamma_\mathrm{M}-\gamma_\sigma)(\gamma_\mathrm{M}+\gamma_\sigma)^2(\gamma_\mathrm{M}-\gamma_{2\overline{3}})(\gamma_{11}\gamma_{21}+8\Omega_\sigma^2)(\gamma_{11}^2\gamma_{12}+8\Gamma\Omega_\sigma^2)^2(\gamma_{31}\gamma_{32}+16\Omega_\sigma^2)}
\end{equation}

\begin{equation}
  \label{eq:Wed4Aug181821CEST2021}
  \mathcal{G}_6\equiv\mathcal{G}_5\ \text{with } \gamma_\mathrm{M}\leftrightarrow-\gamma_\mathrm{M}
\end{equation}

\begin{equation}
  \label{eq:Wed4Aug144101CEST2021}
  \mathcal{G}_7\equiv
  {\begin{multlined}
    32\Gamma^2\gamma_{11}(\gamma_\sigma^2+8\Omega_\sigma^2)\Big[(\gamma_{11}\gamma_{12}+16\Omega_\sigma^2)(\gamma_{11}\gamma_{12}\gamma_{21}^2\gamma_{31}^2\gamma_{32}\gamma_{1\overline{1}}\gamma_{1\overline{2}}\gamma_{2\overline{1}})+{}\\
    \hskip1cm {}+8\gamma_{31}\Omega_\sigma^2\{142\Gamma^7+239\Gamma^6\gamma_\sigma-241\Gamma^5\gamma_\sigma^2-677\Gamma^4\gamma_\sigma^3+77\Gamma^3\gamma_\sigma^4+832\Gamma^2\gamma_\sigma^5+580^\Gamma\gamma_\sigma^6+128\gamma_\sigma^7\}+{}\\
    +64\Omega_\sigma^4\{219\Gamma^6+386\Gamma^5\gamma_\sigma+565\Gamma^4\gamma_\sigma^2+344\Gamma^3\gamma_\sigma^3-98\Gamma^2\gamma_\sigma^4-208\Gamma\gamma_\sigma^5-56\gamma_\sigma^6\}+{}\\
    +1024\Omega_\sigma^6\{15\Gamma^4-11\Gamma^3\gamma_\sigma-4\Gamma^2\gamma_\sigma^2-16\Gamma\gamma_\sigma^3-8\gamma_\sigma^4\}-16384\Omega_\sigma^8\gamma_\sigma\gamma_{21}\Big]
  \end{multlined}
  \over
  \gamma_{21}\gamma_{31}\gamma_{1\overline{1}}(\gamma_\mathrm{M}^2-\gamma_{3\overline{2}}^2)(\gamma_\mathrm{M}^2-\gamma_{4\overline{3}}^2)(\gamma_{11}\gamma_{21}+8\Omega_\sigma^2)(\gamma_{11}\gamma_{12}+8\Omega_\sigma^2)(\gamma_{31}\gamma_{32}+16\Omega_\sigma^2)
  }
\end{equation}


While these expressions are not particularly enlightening, they
provide the most general and exact closed-form formula for the
filtered coherently-driven 2LS. One cannot but marvel at how
Mathematics bring in unexpected factors, 7, 17, 20, to thwart
cancellations of these expression so as to ultimately provide what we
interpretate in physical terms, such as an elbow in a curve that
correspond to the Mollow triplet splitting into three spectral lines.

\bibliography{sci,Books,arXiv} 

\end{document}